\documentclass[amssymb,amsmath,pra,floatfix,twocolumn]{revtex4}
\usepackage{graphicx}
\usepackage{color}
\usepackage{graphicx}
\usepackage{amssymb}
\usepackage{amsmath}
\usepackage{xypic}
\usepackage{array}
\input{Qcircuit}

\begin{document}
\title{Performing Quantum Computing Experiments in the Cloud}

\author{Simon J. Devitt}
\affiliation{Center for Emergent Matter Science, RIKEN, Wakoshi, Saitama 315-0198, Japan.}
\date{\today}

\begin{abstract}
Quantum computing technology has reached a second renaissance in the past five years.  Increased interest from both the private and public sector combined with extraordinary theoretical and experimental progress has solidified this technology as a major advancement in the 21st century.  As anticipated by many, the first realisation of quantum computing technology would occur over the cloud, with users logging onto dedicated hardware over the classical internet.  Recently IBM has released the {\em Quantum Experience} which allows users to access a five qubit quantum processor.  In this paper we take advantage of this online 
availability of actual quantum hardware and present four quantum information experiments that have never been demonstrated before.  We utilise the IBM chip to realise protocols in Quantum Error Correction, Quantum Arithmetic, Quantum graph theory and Fault-tolerant quantum computation, by accessing the device remotely through the cloud.  While the results are subject to significant noise, the correct results are returned from the chip.  This demonstrates the power of experimental groups opening up their technology to a wider audience and will hopefully allow for the next stage development in quantum information technology.
\end{abstract}
\maketitle

\section{Introduction}
The accelerated progress of quantum information technology in the past five years has resulted in a substantial increase in investment and development of active quantum technology.  As expected, access to the first prototypes for small quantum computers have occurred over the cloud, with hardware groups opening up access to their hardware through the classical internet.  The first case was the Center for Quantum Photonics (CQP) at the university of Bristol that connected a two-qubit optical chip to the internet and invited people to design and test their own experiments \cite{QC16}.  This was a remarkable achievement at the time, but suffered from the disadvantage that there was very few experiments in quantum computing that could occur with their cloud based hardware.

Recently IBM did the same thing, and allowed access to a five qubit superconducting chip to the internet community through an interactive platform called the {\em Quantum Experience} (QE) \cite{QE16}.  This approach is substantially more advanced as it allowes access to a five qubit, reprogrammable device and allowed for circuit design, simulation, testing and actual execution of an algorithm on a physical device.  The size of the IBM chip now allows for demonstration of quantum protocols out of the reach of people not associated with advanced experimental groups.  It has already been used by researchers to violate a more general version of Bells inequalities \cite{Al16}.

In this paper we present four separate quantum protocol experiments, designed and executed independently of the IBM team.  We treat the QE website, interface and chip as essentially a black box and run experiments related to four main areas of quantum information; Error Correction \cite{FMMC12}, Quantum Arithmetic \cite{D00}, Quantum Graph Theory \cite{HEB04} and Fault-Tolerant circuit design \cite{PPND15}.  We detail the motivation and design of each experiment, restrict our analysis to the simple output coming from the IBM chip and show that each individual protocol produces valid answers (at low fidelity).  This work will hopefully motivate more people to get involved in cloud based quantum interfaces and encourage experimental groups to open up their hardware for interaction with the general public to increase innovation and development of a quantum technology sector.  
stabilised
Treating the interface as a black box, we designed, simulated and implemented protocols over an array of five qubits.  
The primary results and explanations are provided in the main body of the manuscript, with raw data available in the Supplementary material. 
In each experiment we illustrate simulations provided by the QE interface and the resultant experimental data from the five qubit chip.  In each case we observe results consistent with the theory and the intended output of each protocol.

\section{Results}

{\bf Error Corrected Rabi Oscillations.} The first experiment implemented on the IBM chip is a basic Rabi oscillation spectra across a {\em logically} encoded qubit using a distance two surface code.  Surface code quantum computing \cite{FMMC12}, is now the standard model used for large-scale quantum computing development \cite{DFSG08,DMN08,JMFMKLY10,YJG10,NTDS13,MRRBMD14,AMK15,LWFMDWH15,LHMB15,HPHHFRSH15,ONRMB16} and results from superconducting and other technologies show extraordinary promise \cite{B14,C15,BKM16,DJ13,SSM16}.  While the distance three, 5-qubit code \cite{PhysRevLett.77.198} could be used with the IBM QE, we focus on the surface code due to its relevance to larger quantum architectures. The five qubit surface code has already been investigated by the IBM team themselves \cite{C15}, but in this experiment, the goal is not to artificially inject errors into the system, but rather to conduct a standard protocol used in the initial demonstration of a two-level controllable quantum system and see if an error corrected version of the protocol shows some advantage.  

\begin{figure}
\begin{center}
\resizebox{\linewidth}{!}{\includegraphics{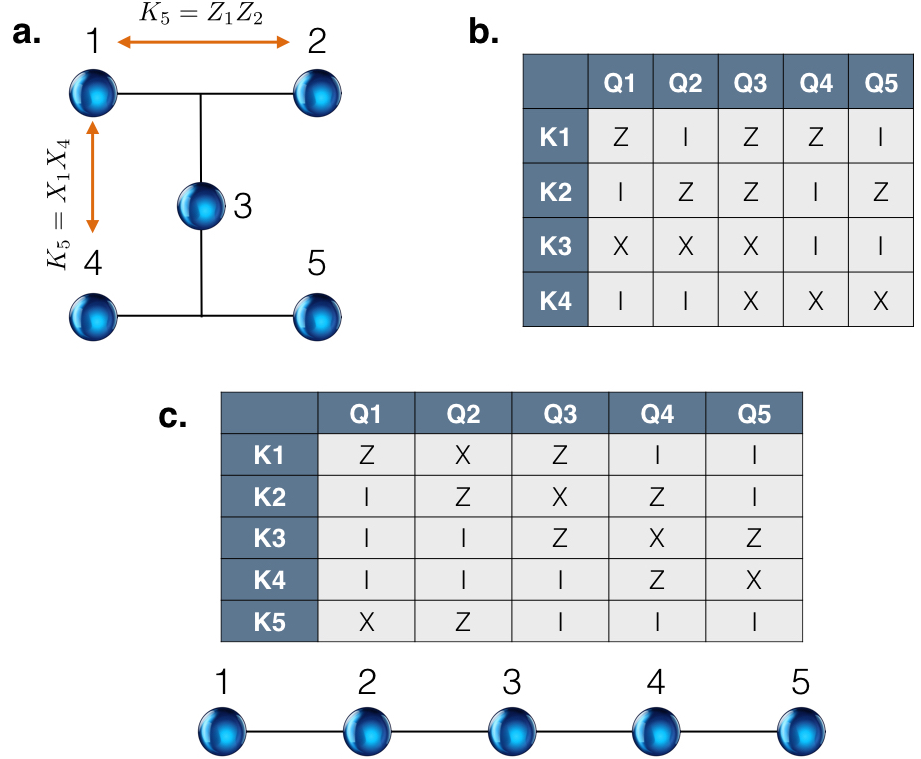}}
\end{center}
\caption{{\bf (colour online) Distance two surface code.}  Figure a.) illustrates the code structure, with two plaquette and two vertex stabilisers.  Logical operators run left to right ($Z_L$) or top to bottom ($X_L$).  Figure b. is the stabiliser set for the distance two surface code.  Figure c. is the structure and stabiliser set for a linear cluster state.  Once a Hadamard is performed on qubits 1,3,5, the state is equivalent to a distance two surface code initialised in the $\ket{0}_L$ state.}
\label{fig:stab}
\end{figure}

A distance two surface code consists of an array of five qubits, illustrated in Figure \ref{fig:stab}a) stabilised by the four operators in Figure \ref{fig:stab}b).
With a fifth stabiliser $K_5 = Z_1Z_2$ ($K_5 = X_1X_4$) used to specify the logical $Z$ ($X$) state of the encoded qubit.  
The $Z$-stabilisers ($\{K_1,K_2\}$) are used to detect bit flip errors while the $X$-stabilisers ($K_3,K_4$) are used to correct phase errors.  Being distance two, this code can only detect errors (there is insufficient information to determine location of any detected bit or phase error, with certainty).  

Initialising a distance two surface code in the $\ket{0}_L$ state is equivalent to preparing a five qubit linear cluster.  The five qubit linear cluster has stabiliser set illustrated in Figure \ref{fig:stab}c). After a Hadamard is applied to qubits one, three and five we can identify $K'_5$ as the $+1$-eigenstate of the logical $Z$ operator and after multiplying stabilisers $K'_2K'_5 = K_1$ and $K'_2K'_4 = K_2$ we regain the stabiliser set for the distance two surface code.  

\begin{figure}
\begin{center}
\resizebox{\linewidth}{!}{\includegraphics{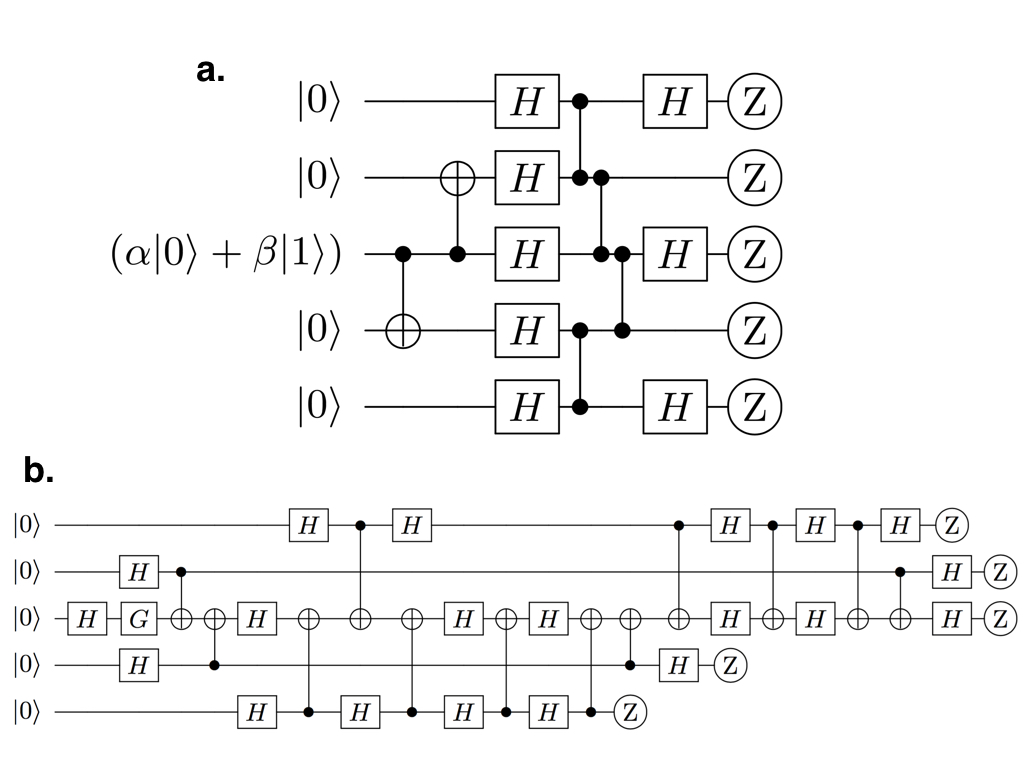}}
\end{center}
\caption{{\bf Circuits to encode arbitrary states into the surface code}.  Figure a) is the general circuit, where the state 
$\alpha\ket{+++}_{1,3,5}+\beta\ket{---}_{1,3,5}$ (where 1,3,5 are the central three qubits in the QE interface at input) is first prepared and then a linear cluster state is created afterwards.  Following the last three Hadamard gates, the logic state $\ket{0}_L + \ket{1}_L$ is prepared. Figure b) is the circuit implementation in the QE 
interface.  Only the third qubit in the IBM chip can be coupled to via a CNOT (and it always acts as the target qubit), hence various 
SWAP gates (decomposed into CNOTs and Hadamards \cite{NC00} are used).  The gate $G$ is the set of successive $T$ rotations 
used to sequentially rotate the qubit into the state, $e^{i n(\pi/8)X_L}\ket{0}$ for $n \in \{0,..,7\}$.  After the state is prepared it is immediately measured in the $Z$-basis.  After the circuit is run, qubit ordering is (4,1,2,5,3).}
\label{fig:circuit}
\end{figure}

To perform an error correction version of a Rabi oscillation experiment, we instead require the ability to encode a rotated logical state of the form $e^{i\theta X_L}\ket{0_L}$ which cannot be directly prepared in the encoded space.  Therefore we prepare a single qubit in the rotated state and encode.  This is not a fault-tolerant operation and the fidelity of this initial rotated state will bound the error on the final state (even if every other gate in the circuit was perfect).  The direct encoding is shown in Figure \ref{fig:circuit}a), where we prepare the state $\alpha\ket{+++}_{1,3,5} + \beta\ket{---}_{1,3,5}$ and then create the linear cluster state from this initial state.  This places the surface code into the state $\alpha\ket{0}_L + \beta\ket{1}_L$.  

As with a single qubit Rabi experiment, we measure the probability of measuring a $\ket{0}_L$ state after rotating the input with successively larger angles, $\theta$.  In the QE interface, only Clifford $+T$ gates are available, and constructing rotations smaller than $R_x(\pi/4)$ would require decompositions into Clifford $+T$ sets \cite{RS14,GKMR14}.  But for this experiment we instead use successive $\pi/4$ rotations ($T$-gates) around the Bloch sphere where an eight gate cycle will oscillate the state from $\ket{0}_L \rightarrow \ket{1}_L$ and back to $\ket{0}_L$.  

As the QE chip only allows for CNOT gates to operate with the central qubit in their star geometry, additional SWAP operations are needed, which are also decomposed into CNOT and Hadamard.  The full circuit implemented is illustrated in \ref{fig:circuit}b), Where the gate $G$ is each successive step in the 
rotation, $G\in \{T^0,...,T^8\}$.  The input ordering of qubits in the QE interface is (2,1,3,5,4) and at output the ordering is (4,1,2,5,3).

Once the state is prepared, we can measure in the $Z$-basis on each of the five qubits and calculate the parity of $K_5$ to determine the logical state.  Since when measuring in the $Z$-basis, results are modified by the presence of bit flip errors, we can use the parity of the $Z$-stabilisers, to post-select from the 8192 runs allowed by the QE interface only instances where $K_1$ and $K_2$ return even parity results.  
The raw data is included in the supplementary material.

\begin{figure}
\begin{center}
\resizebox{\linewidth}{!}{\includegraphics{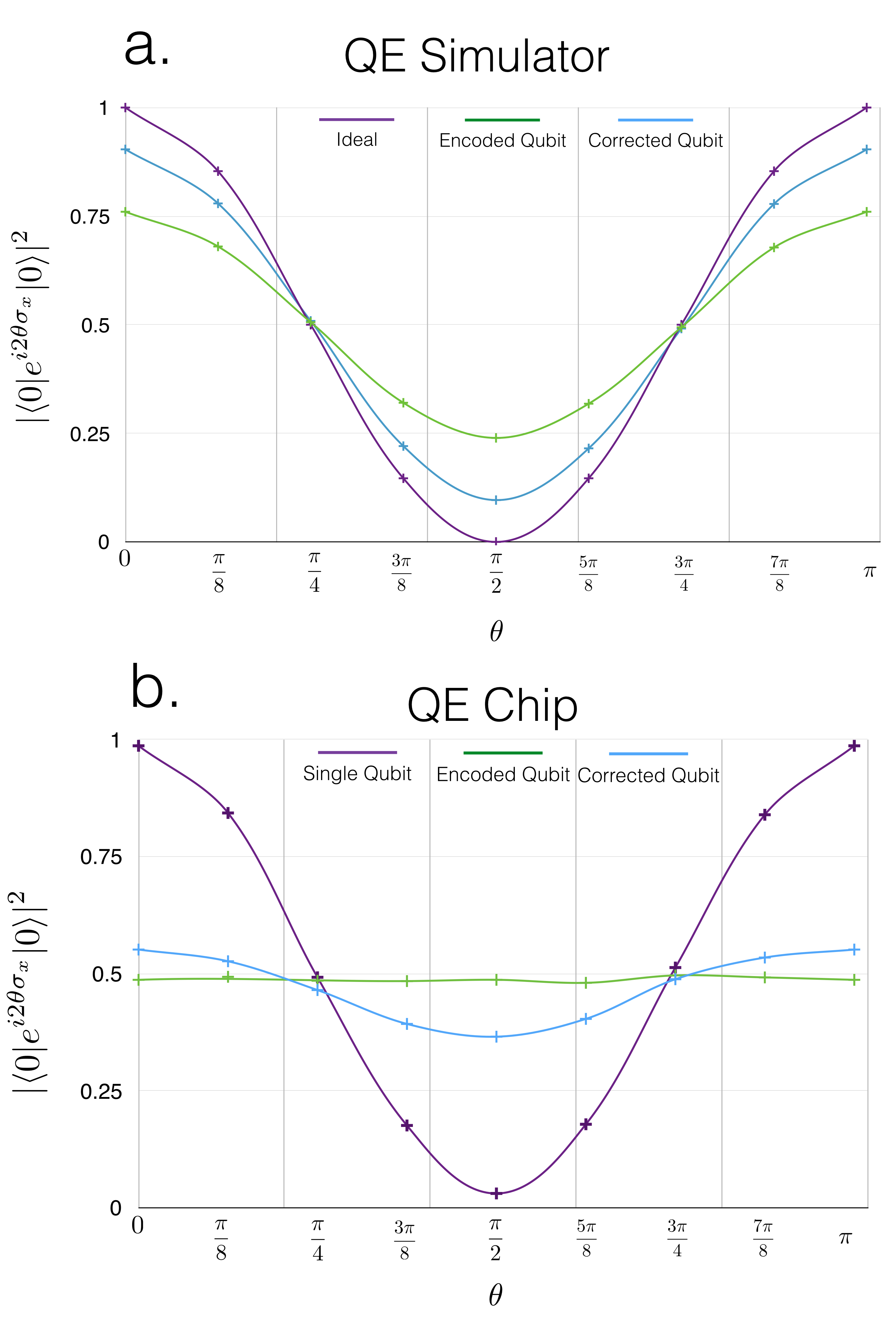}}
\end{center}
\caption{{\bf (colour online) Simulation and Experimental result from the QE device.}  In Figure a) we have simulated plots for the ideal Rabi curve (purple), the encoded but {\em un-corrected} code (green) and the post-selected, error-corrected results (blue).  An improvement when correction is applied is clearly seen.  In Figure b) we have the results directly from the QE chip.  For the {\em corrected} state, a low visibility oscillation is observed.  The purple curve now represents successive rotations on an {\em individual} qubit and so has high visibility (but still is not perfect).  Each data point was obtained using the maximum sampling in the QE interface of 8192 shots, and a sampling error for each point (omitted) can be calculated as SE $=\sqrt{P_c(1-P_c)/8192}$, where $P_c$ is the probability of each measurement.} 
\label{fig:surfaceexp}
\end{figure}

Once the state is prepared we first simulated the expected output using the master equation solver included with the QE interface and the simulation results are illustrated in Figure \ref{fig:surfaceexp}a).  On this plot there is three curves.  The first is the theoretical optimal, where the Rabi oscillations follow a $\cos^2(\theta)$ function.  The second curve, with the lowest visibility, is when we run the circuit in Figure \ref{fig:circuit} but {\em do not} post-select on trivial syndrome results for $K_1$ and $K_2$.  The visibility in this case is clearly lower than the ideal case, due to errors accumulating in the larger circuit.  The third curve, which sits between the two, is where post-selection occurs and we are discarding any results where a non-trivial syndrome is detected.  In simulations, this gives clearly better performance than the non-error-corrected case, but is still far from ideal.  

Illustrated in Figure \ref{fig:surfaceexp}b) is the actual experimental data returned from the QE interface.  Each data point (representing successive rotations of the input state between $\ket{0}_L$ and $\ket{0}_L$ and the rotation angle is $2\theta$, such that a rotation from $\ket{0}_L \rightarrow \ket{0}\rangle$ occurs for $\theta=\pi$.) was taken using the maximum sampling instances of the QE chip (8192).  For the purple illustrated in Figure \ref{fig:surfaceexp}b), we performed a Rabi experiment using a bare single qubit (via repeated rotations by $T$ and then immediate measurement), this showed unsurprisingly the highest visibility.  The data when we encoded into the surface code and did not perform the corrective operations, is essentially random (showing a 50\% probability of measuring $\ket{0}_L$ or $\ket{1}_L$).  Once the data is post-selected on trivial syndrome results, we again see a non-zero visibility, indicating that the error correction properties of the code are helping to restore the oscillation.  The standard error can be calculated as SE $= \sqrt{P_c(1-P_c)/8192}$, where $P_c$ is the probability of a given measurement result.  The SE is not illustrated in Figure \ref{fig:surfaceexp}. 
%but is included in supplementary material.

The complexity of the circuit to perform the encoding and the fact that the QE chip does not have sufficient control or accuracy to perform a fault-tolerant experiment, means that the error-corrected version {\em does not outperform} a non-error corrected version of the same experiment.  But active decoding of the information does result in a better results when compared to the same circuit without active decoding.  Additionally, the simulator bundled with the QE interface overestimates the performance of the chip, compared with the actual experimental data.

%Additional data and experiments are shown in the supplementary material that replicates the results approximately a week later so show the stability of the QE chip and a similar experiment performed in the $\ket{+,-}_L$ basis which produces the same results but for correction of $Z$-errors. 

{\bf Fourier Addition}
Our second experiment is programmable quantum arithmetic \cite{D00}.  We design a circuit that performs addition between two quantum registers, $\ket{a}$ and $\ket{b}$ by performing a Fourier transform on register $\ket{a}\otimes\ket{b} \rightarrow \ket{a}\otimes \psi(\ket{b})$, controlled phase rotations taking $\ket{a}\otimes\psi(\ket{b})\rightarrow \ket{a}\otimes\psi(\ket{a}+\ket{b})$ and an inverse Fourier transform, giving, $\ket{a}\otimes\psi(\ket{a+b}) \rightarrow \ket{a}\otimes \ket{a+b}$.  The addition naturally occurs modulo 2 as the registers $\ket{a}$ and $\ket{b}$ are only each 2-qubit registers.  

Even though the QE system has the ability to operate over five qubits, a Quantum Fourier Transform (QFT) over anymore than two qubits require $\sqrt{T}$ gates, which would need to be approximated via Clifford $+T$ sequences, for which there are insufficient resources.  Two qubit Fourier addition is possible as the smallest rotation needed for circuit decompositions are $T$-gates, available in the QE interface.

\begin{figure*}[ht!]
\begin{center}
\resizebox{0.9\linewidth}{!}{\includegraphics{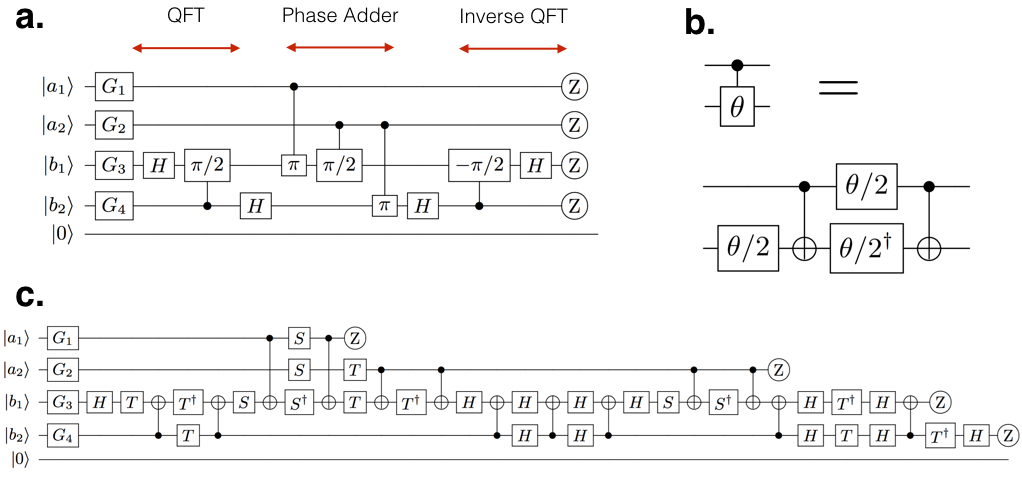}}
\end{center}
\caption{{\bf (colour online) Fourier addition circuits.} In Figure a) we illustrate the Fourier addition circuit on two registers.  First a QFT is performed on the target register, after which controlled phase rotations are used to perform the addition.  An inverse QFT then returns the target register to the computational basis.  The gates $G_i$, $i\in \{1,..,4\}$ are used to program in various inputs used in the simulations.  Figure b) shows controlled rotation decompositions.  For rotations smaller than $\pi/2$, the decomposition requires gates of the form $^n\sqrt{T}$, which are not available in the QE interface without performing 
Clifford $+T$ decompositions \cite{RS14,GKMR14}.  In Figure c) we illustrate the circuit implementation for the QE chip. Note that in Figure c. the target register, $\ket{b_1}\ket{b_2}$ is flipped on output.}
\label{fig:fouriercircuit}
\end{figure*}

In Figure \ref{fig:fouriercircuit}a) we illustrate the addition circuit, the decomposition circuits used for controlled phase rotations [Figure \ref{fig:fouriercircuit}b)] and the actual circuit implemented in the QE interface [Figure \ref{fig:fouriercircuit}c)].  Once again, all CNOTs need to involve the central qubit in the architecture ($Q_3$) as the target, in accordance with the geometry of the QE hardware.  The gates $G_i$, $i \in \{1,..,4\}$ are used to prepare the input states from the initial $\ket{0}$ states in the QE chip, and at the end of the circuit everything is measured in the $Z$-basis.  

\begin{figure*}[ht!]
\begin{center}
\resizebox{0.8\linewidth}{!}{\includegraphics{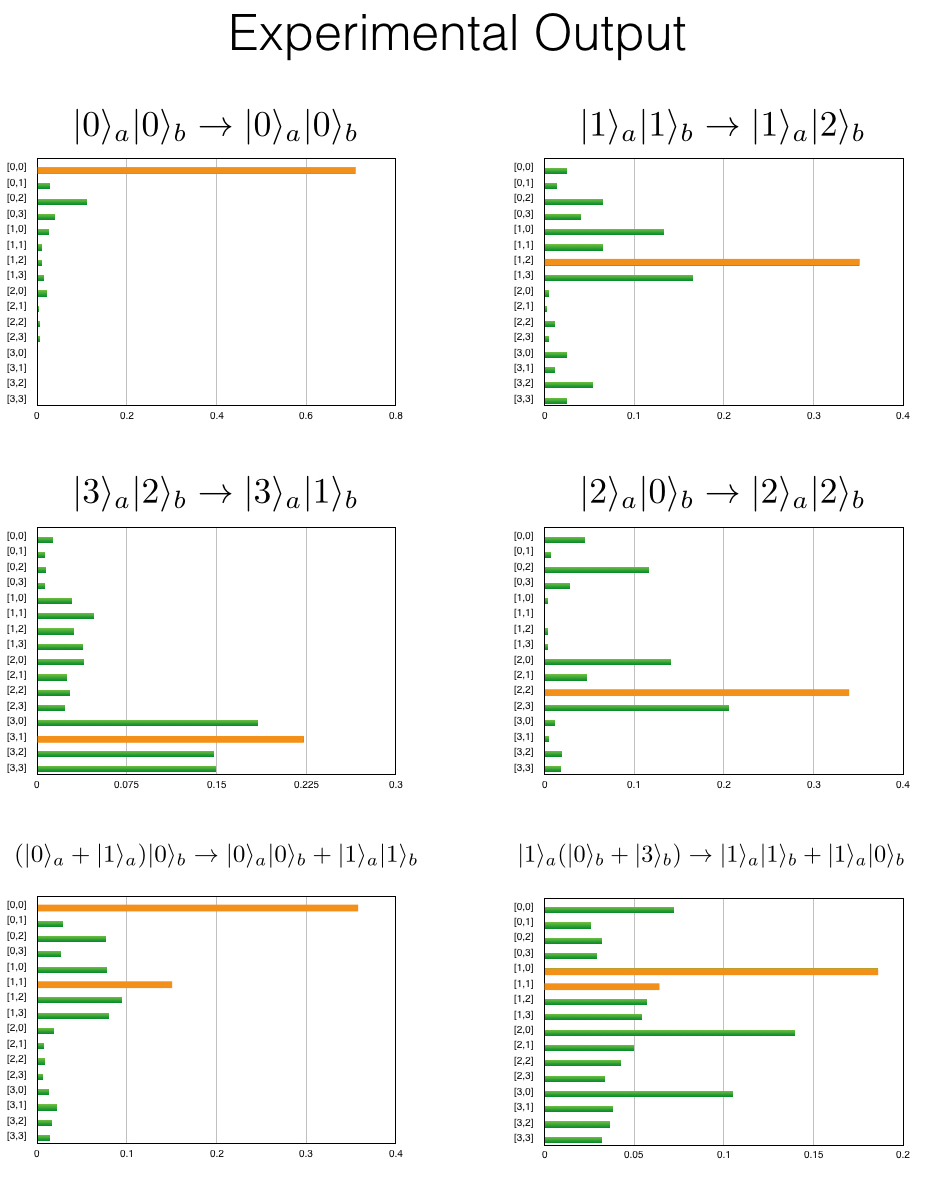}}
\end{center}
\caption{{\bf (colour online) Experimental Fourier addition.}  Each plot shows the 32 possible binary outputs.  The input/output mapping for 
each plot is specified, and the {\em correct} binary outputs are illustrated in Orange on each plot.  For all experiments (except 
where a Bell state is created on the target register), the correct output is observed a plurality of the time.}
\label{fig:fourierexp}
\end{figure*}

\begin{figure*}[ht!]
\begin{center}
\resizebox{0.8\linewidth}{!}{\includegraphics{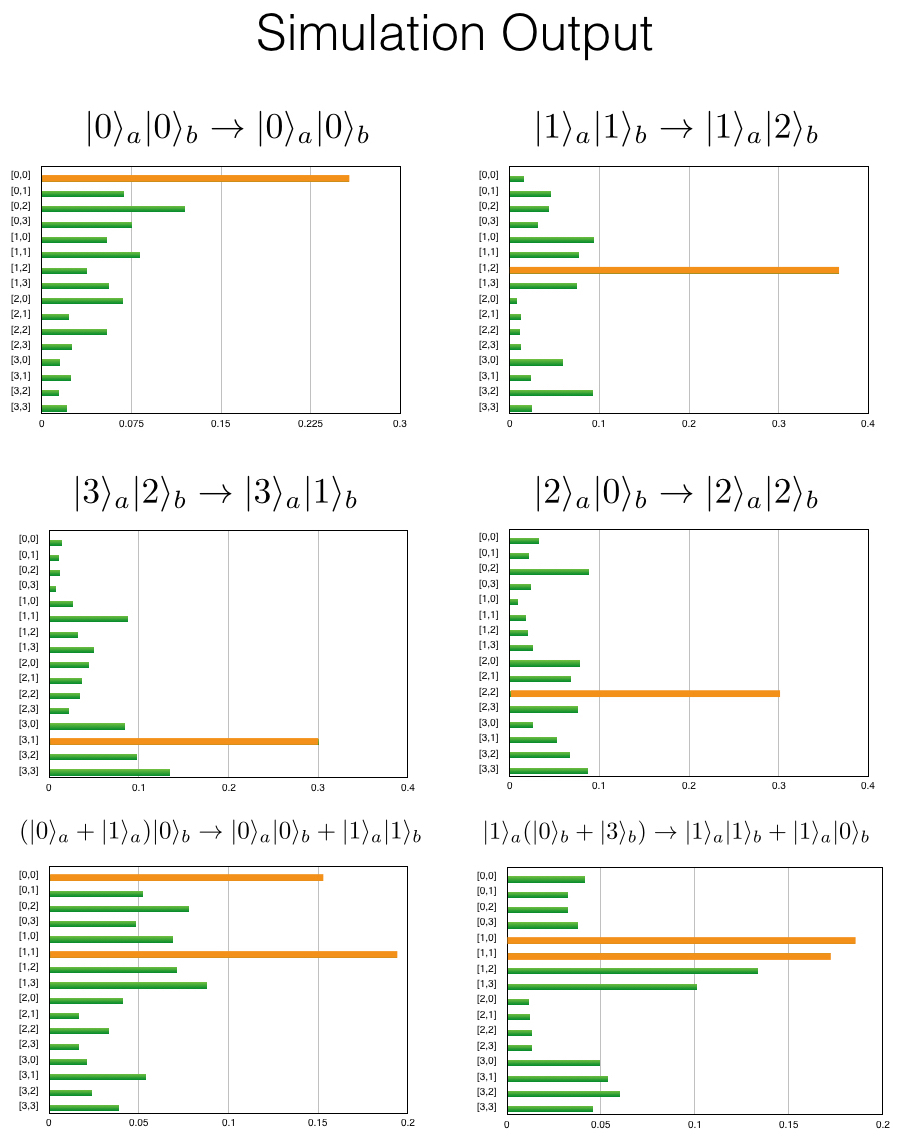}}
\end{center}
\caption{{\bf (colour online) Simulated Fourier Addition.}  We use the QE interface to first simulate the addition circuit.  Results are closer to experimental output than the error-corrected Rabi experiment, and in some cases show more noise than what was output from the chip itself.}
\label{fig:fouriersim}
\end{figure*}

As with the error corrected Rabi oscillation, the final circuit is quite large and hence errors will accumulate and the probability of observing successful output will drop.  We again ran these circuits in both the QE simulator and the actual QE chip for comparison.  Shown in Figure \ref{fig:fouriersim} are the simulation results and Figure \ref{fig:fourierexp} the experimental results for a selection of the 16 possible binary inputs of $\ket{a}$ and $\ket{b}$, along with various superposition and entangled inputs. 
the complete data set is contained in the supplementary material.  
The simulations and experimental results were averaged over 8192 runs and the standard error (SE) omitted from the plots.  The simulations and experiments show similar levels of noise in their outputs.  The difference between the two is not as clear as the error-corrected Rabi oscillations shown earlier. 

The size of the addition circuit is larger than the error corrected Rabi experiment, hence noise in the QE chip is stronger.  However, again we see the {\em right} answer returned a plurality of the time, from the 8192 runs performed.  The only exception is when we used an entangled input on the target register.  Illustrated in Figure \ref{fig:fourierexp} on the bottom right plot, when the transformation should be $(\ket{0100}+\ket{0111})/\sqrt{2} \rightarrow (\ket{0101}+\ket{0100})/\sqrt{2}$.  In this case, the correct answer couldn't be inferred from the experimental output.  

We expect that experimental refinement of the chip itself will lead to less noise accumulation such that desired results are returned with much higher probability.  As experimental error rates go down, these experiments would be useful as a benchmark to ascertain if performance with large circuits increases over time. 

{\bf Local Complementarity in Quantum Graph States}
The third experiment performed with the QE platform is a unique property of quantum graph states called local complementarity \cite{HEB04}.  Local complimentary is where the structure of a quantum graph can be changed via {\em only} local operations.  We demonstrate the creation of a five qubit graph state (by measuring appropriate graph stabilisers and confirming measuring even parity of each operator), we then permute this graph using local operations and confirm, through the measurement of the new stabilisers, that the underlying state has changed.  A minimum of four qubits is required to perform any type of graph complementarity experiment (as all graphs containing two or three qubits have only a single local-equivalence class \cite{HEB04}) and 
five qubits give a large number of non-equivalent classes that we could test.  The first example is creating a GHZ state (star graph) and through local complementarity, perturb this to a completely connected graph and then perturb it back to another star graph with a different qubit acting as the star node.  
The second experiment is running several {\em orbital} steps on a second input graph, showing the creation of each new, locally equivalent graph.

\begin{figure}[ht!]
\begin{center}
\resizebox{\linewidth}{!}{\includegraphics{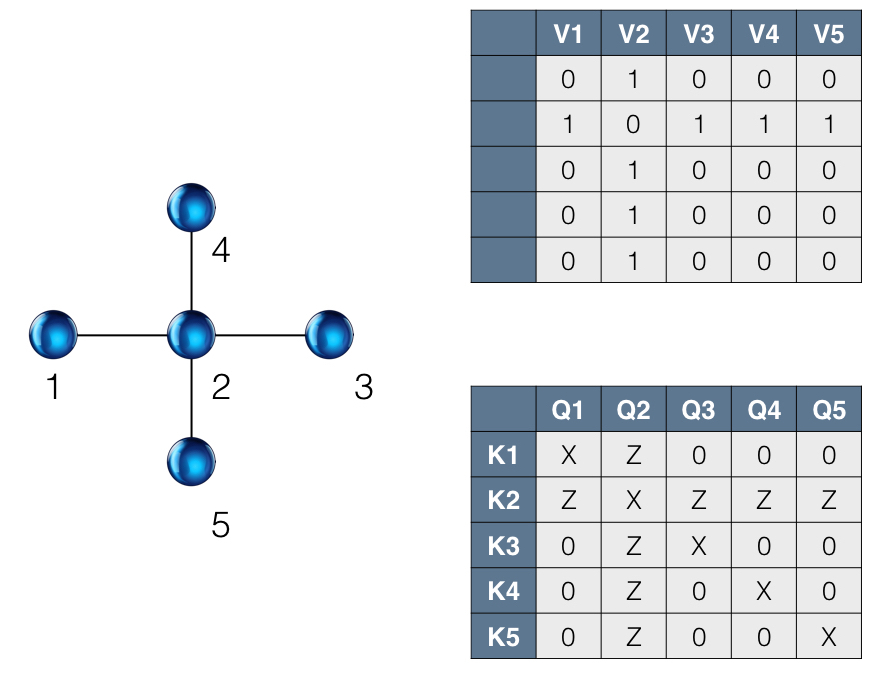}}
\end{center}
\caption{{\bf (colour online) The link between a classical graph and the corresponding quantum graph.}.  The classical star graph can be represented as an adjacency matrix, with ones present to indicate the four bonds from the central node to each leaf.  To convert this to the equivalent quantum graph state, we construct a stabiliser table by replacing each 1 with a $Z$ operator, $X$ operators on the diagonal and $I$ everywhere else.  Each row of the adjacency matrix is now one of the five stabilisers of the quantum star graph.}
\label{fig:star}
\end{figure}

A quantum graph state is easily defined from the classical graph that describes it.  A classical (undirected, finite) graph, $G = (V,E)$, is a set of $N$ nodes, $V\in \{1,...,N\}$ connected by sets of edges, $E \subset V^2$ such that $E_{i,j} = 1$ for any two connected nodes, $V_{i,j}$ and $E_{i,j}=0$ if $V_{i,j}$ are not connected.  To convert this classical graph state to a quantum graph state, we use the adjacency matrix of the classical graph to form a set of stabiliser operators that are used to specify the quantum graph.  The conversion of a star graph to the stabiliser set is illustrated in Figure \ref{fig:star}.   

Once a graph is written, an adjacency matrix can be formed that contain  a 1 for each edge, $E_{i,j} = 1$ and 0 everywhere else.  To convert an adjacency matrix into a list of stabilisers for the quantum graph state requires taking each row of the adjacency matrix, replace the 0 on the diagonal with the Pauli $X$-operator, any 1 entry with the Pauli $Z$-operator and the Identity matrix everywhere else.  For an $N$-qubit graph, there will be $N$-operators corresponding to the $N$-rows of the classical adjacency matrix.  Creating a quantum state simultaneously stabilised by these $N$-operators will produce the required quantum state.  

\begin{figure}[ht!]
\begin{center}
\resizebox{\linewidth}{!}{\includegraphics{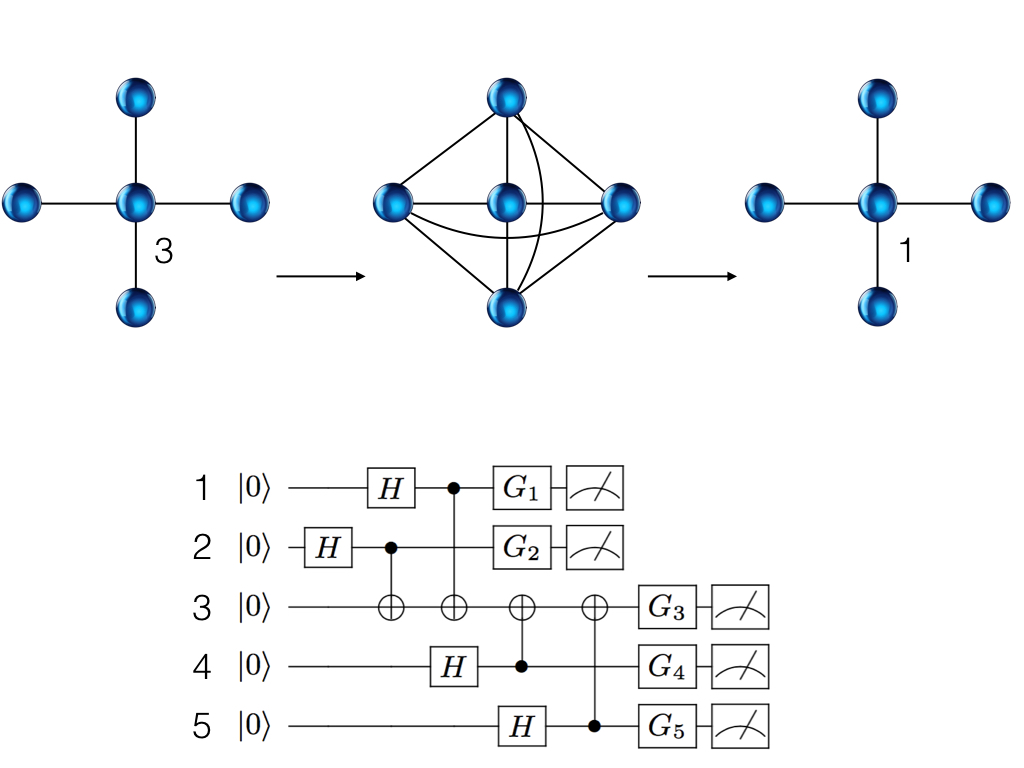}}
\end{center}
\caption{{\bf (colour online) circuits to implement graph complementarity and stabiliser measurements.}.  In the experiment we initially prepare a 
five qubit star graph, use local complementarity to convert this to a completely connected graph and then permute again to reconvert to a star graph with a different qubit acting as the central node.  After each conversion we measure the five associated stabilisers.  Illustrated is the quantum circuit necessary to do this in the QE interface.  The initial part of the circuit creates the star graph.  The gates $G_i$, $i \in \{1,..,5\}$ allow us to perform the graph complementarity operations (with $S$- and $T$-gates) and/or to measure in either the $X$ or $Z$-basis.}
\label{fig:graph}
\end{figure}

Local complementarity operators are quite simple, and consist of choosing one of the qubits of the quantum graph, denoted the {\em node}, applying a $\sqrt{X} \equiv HSH$ operator to that qubit, and a $\sqrt{Z} \equiv S$ operator to any qubit it is connected to ({\em leaf}).  This has the effect of creating an edge directly between the leaves or deleting an edge if it originally existed.  These local complementarity rules create what is known as an {\em orbital} and forms a closed set of locally equivalent quantum graphs.  Different orbital groups are not locally equivalent to each other and the number of different orbital groups grows as the total number of qubits increase.  Our first experiment is illustrated graphically in Figure \ref{fig:graph} along with its required quantum circuit.  

\begin{figure*}[ht!]
\begin{center}
\resizebox{0.8\linewidth}{!}{\includegraphics{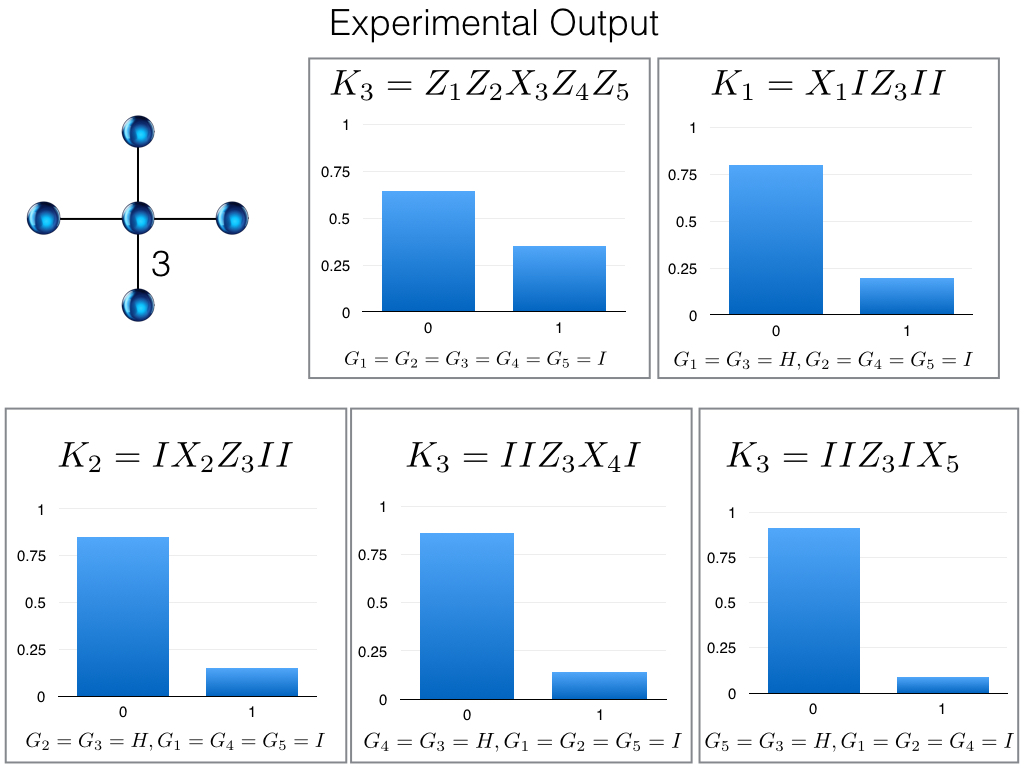}}
\end{center}
\caption{{\em (colour online) Experimental results for the initial star graph.}  For each plot, we measure the parity of the associated stabiliser.  The settings for each of the $G_{i\in 5}$ gates are illustrated for each.}
\label{fig:graph1}
\end{figure*}

\begin{figure*}[ht!]
\begin{center}
\resizebox{0.8\linewidth}{!}{\includegraphics{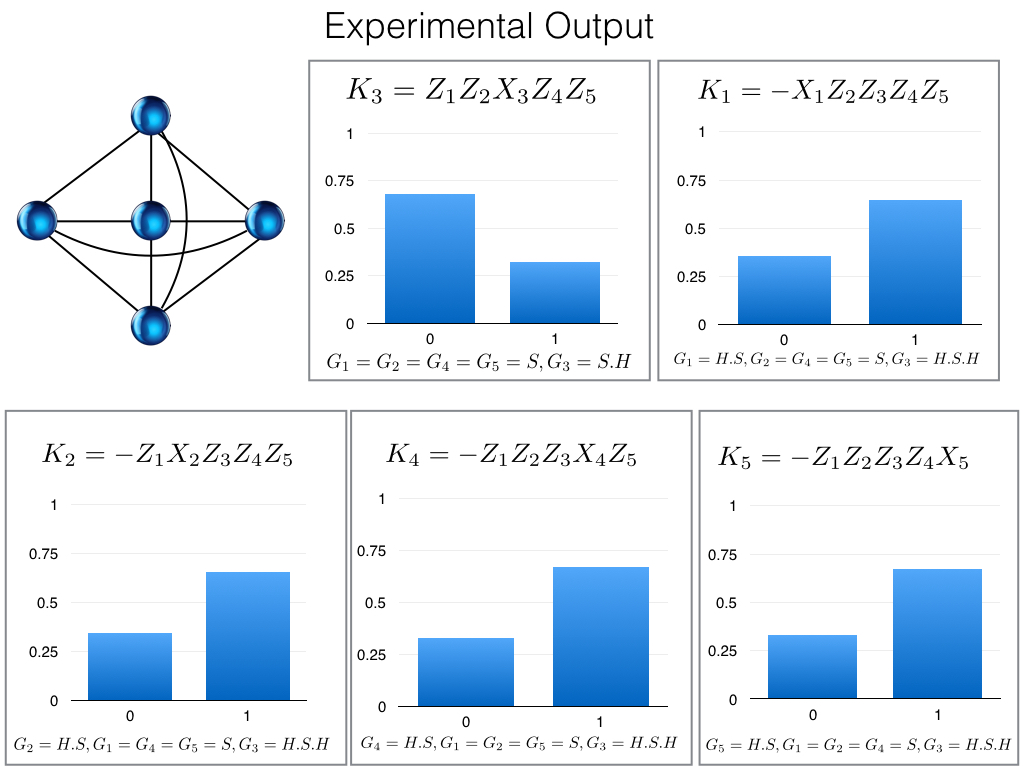}}
\end{center}
\caption{{\bf (colour online) Experimental results for the completely connected graph.}  The completely connected graph is constructed by performing a $HSH$ gate on qubit one and an $S$ gate on every other qubit.  The permutation of the stabiliser set after this operation can cause some eigenvalues to flip between $+1$ and $-1$.  For each plot, we measure the parity of the associated stabiliser.  The settings for each of the $G_{i\in 5}$ gates are illustrated for each.}
\label{fig:graph2}
\end{figure*}

\begin{figure*}[ht!]
\begin{center}
\resizebox{0.8\linewidth}{!}{\includegraphics{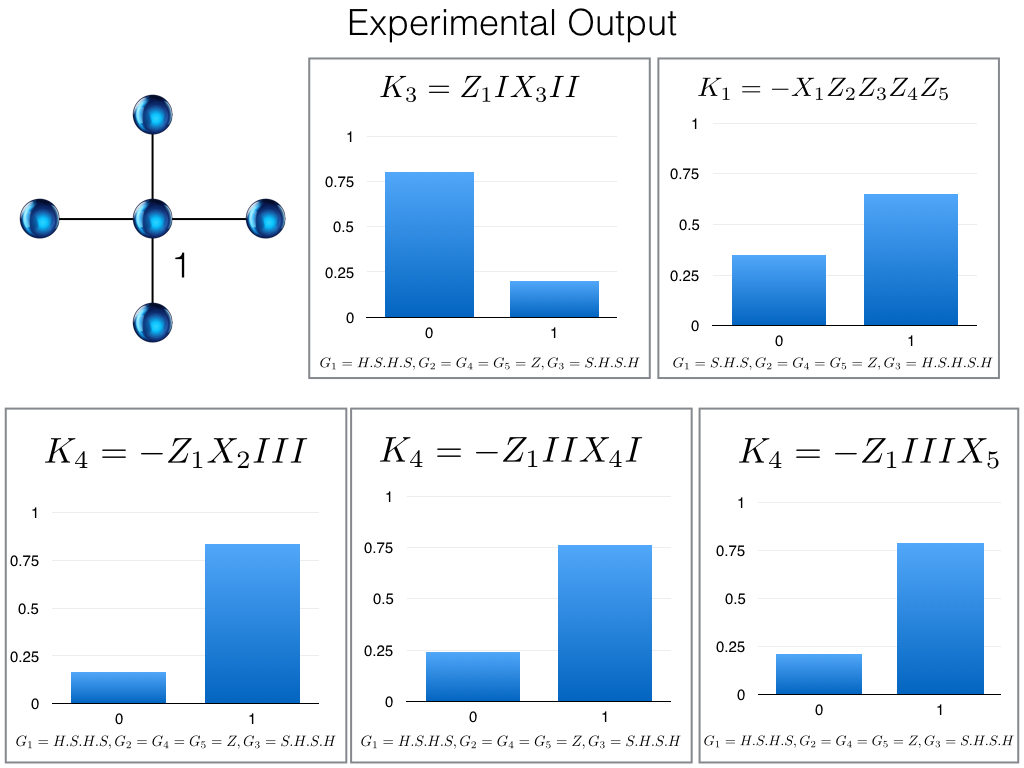}}
\end{center}
\caption{{\bf (colour online) Experimental results for the final star graph.}  We construct the final star graph (with qubit one acting as the node) by performing $HSH$ to qubit one and $S$ to every other qubit.  This permutation does not change the eigenvalue of any of the stabilisers.  For each plot, we measure the parity of the associated stabiliser.  The settings for each of the $G_{i\in 5}$ gates are illustrated for each.}
\label{fig:graph3}
\end{figure*}

\begin{figure*}[ht!]s
\begin{center}
\resizebox{0.8\linewidth}{!}{\includegraphics{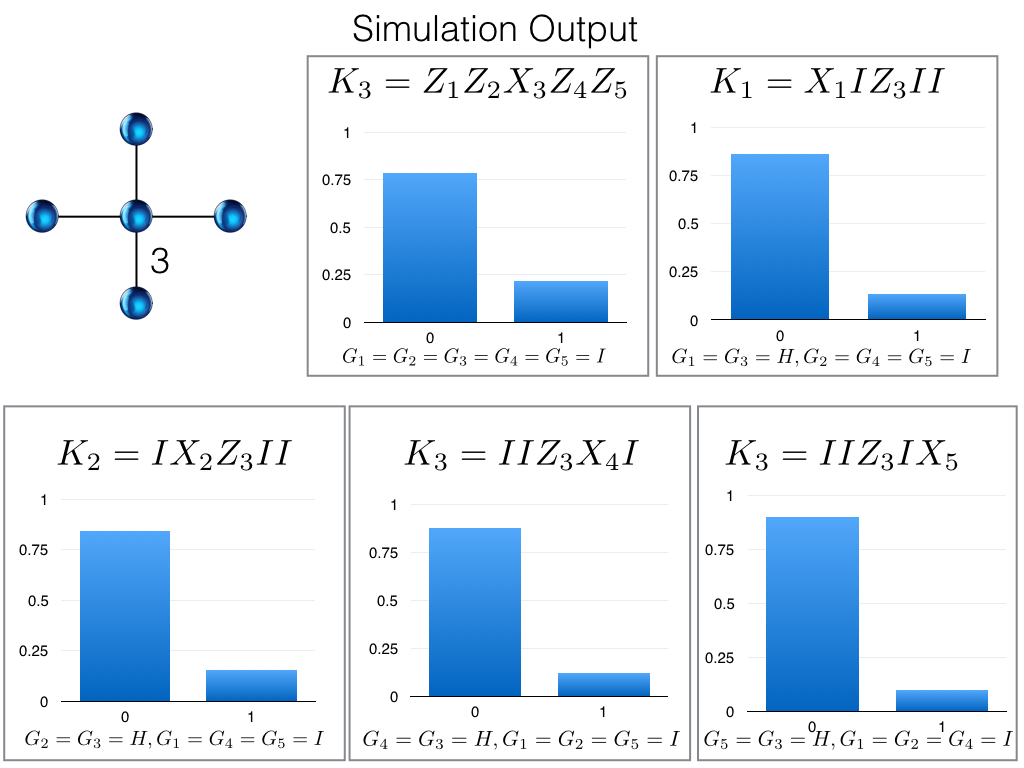}}
\end{center}
\caption{{\bf (colour online) Simulated results for Figure. \ref{fig:graph1}.}}
\label{fig:compsim1}
\end{figure*}

\begin{figure*}[ht!]
\begin{center}
\resizebox{0.8\linewidth}{!}{\includegraphics{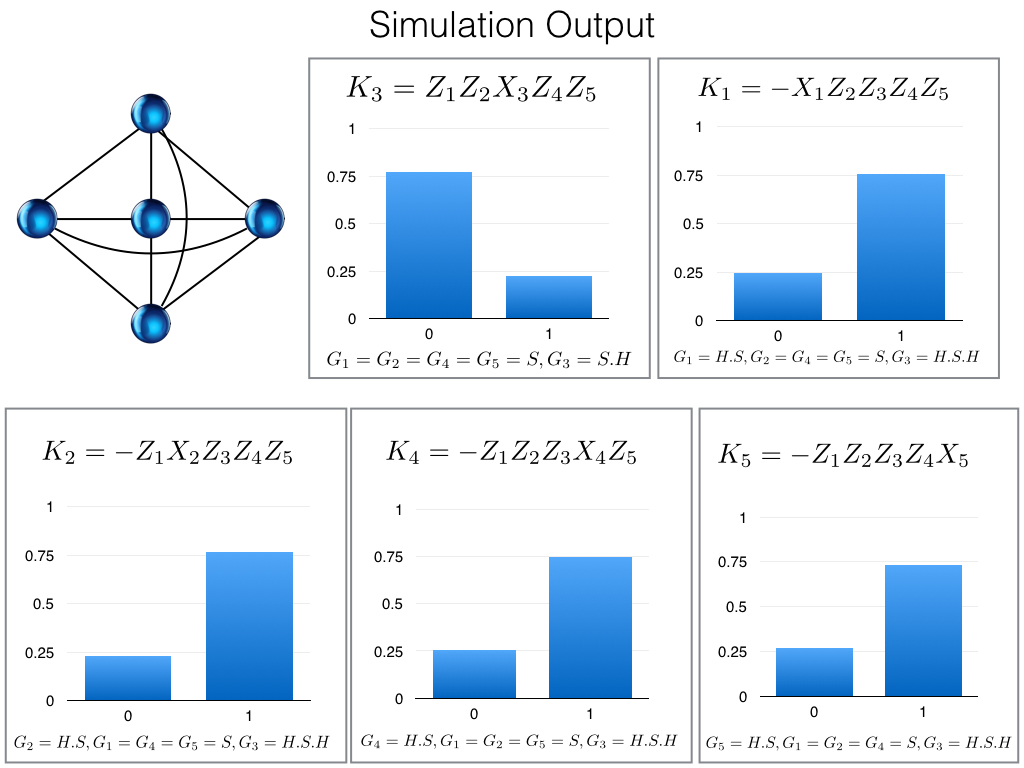}}
\end{center}
\caption{{\bf (colour online) Simulated results for Figure. \ref{fig:graph2}.}}
\label{fig:compsim2}
\end{figure*}

\begin{figure*}[ht!]
\begin{center}
\resizebox{0.8\linewidth}{!}{\includegraphics{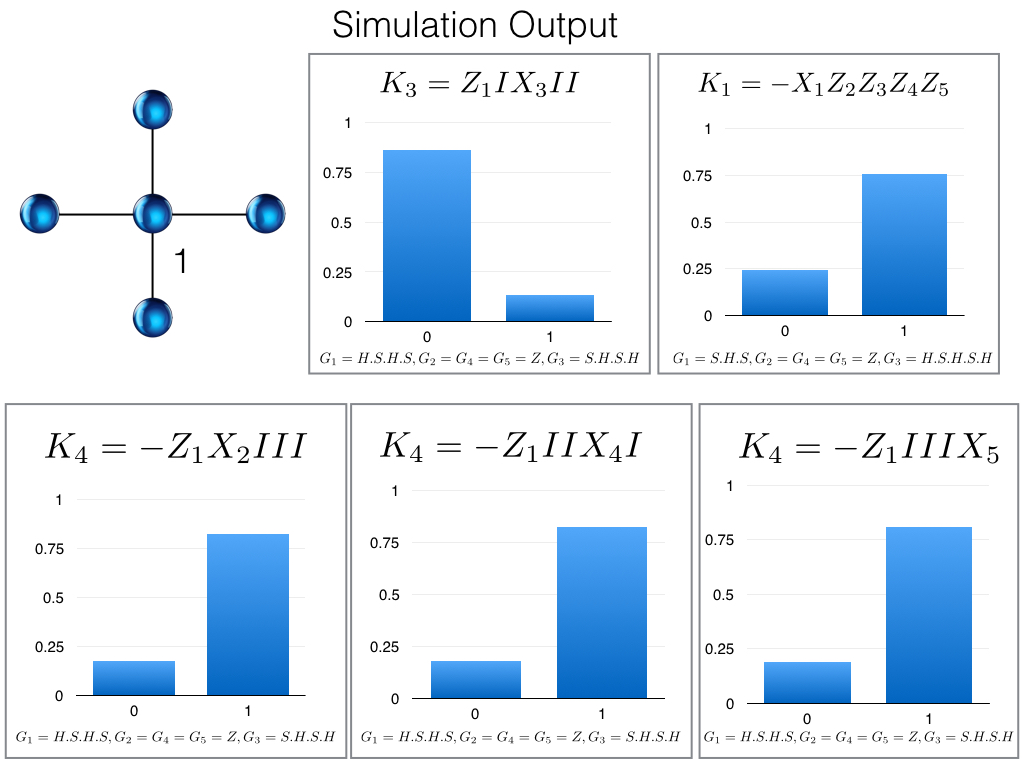}}
\end{center}
\caption{{\bf (colour online) Simulated results for Figure. \ref{fig:graph3}.}}
\label{fig:compsim3}
\end{figure*}

We first create a 5-qubit star graph that is stabilised by the operators in Figure \ref{fig:star}, and then through the local gates $G_i$, $i\in \{1,..,5\}$ we are either measuring the stabiliser of the star graph or through local complementarity we first convert the graph and measure the parity of the stabilisers associated with the new graph.  The experimental results from the QE chip is illustrated in Figures \ref{fig:graph}-\ref{fig:graph3}, and the simulated results are illustrated in 
Figures \ref{fig:compsim1}-\ref{fig:compsim3}.  Again, each plot was generated using 8192 experimental runs, and the SE is omitted from each plot.



As expected, the calculated parities for each stabiliser match what is measured.  Some parities flip during the graph complementarity operations, but these flips could be reversed by applying appropriate Pauli operations.  As the circuit is quite small compared to other experiments, the degree to which we see nearly 100\% probability of the correct parity is high.  This experiment showed how to use local complementarity to permute the structure of a quantum graph.  In this case we went from a star-graph with qubit 3 acting as the node, to a completely connected graph and back to a star-graph with now qubit one acting as the node.  
Another example of five qubit complimentary is shown in the supplementary material.

This type of experiment for all non-equivalent five qubit graphs is fairly simple to perform in the QE interface.  Preparing any five qubit graph state does not require as many gates as either error-corrected Rabi oscillations {\em or} Fourier addition and hence the QE chip should produce the correct state (with reasonably high fidelity).  Investigating the structure of these locally-equivalent graphs and how they can be utilised as communications links has the potential to open up an area of quantum network analysis that is not available in the classical world.

\begin{figure}[ht!]
\begin{center}
\resizebox{\linewidth}{!}{\includegraphics{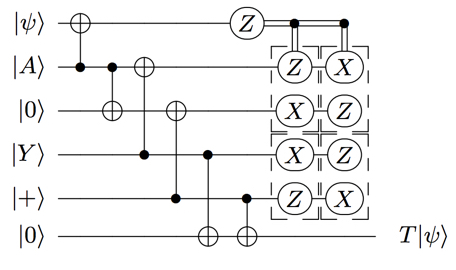}}
\end{center}
\caption{{\bf Teleported $T$-gate with deterministic circuitry}.  Built using selective source and destination circuits \cite{F12+}, 
this circuit will first apply the $T$-gate, through a teleportation circuit with a magic state, $\ket{A} = \ket{0}+e^{i\pi/4}\ket{1}$, ancilla.  
A possible $S$-correction may be needed, requiring the second magic state ancilla, $\ket{Y} = \ket{0} + i\ket{1}$.  By selecting one of two-choices of 
$X$- and $Z$-basis measurements (based on the top $Z$ measurement), the $S$ correction is applied or not.  Hence the output 
is always the $T$-gate operating on the input state, $\ket{\psi}$.}
\label{fig:Tgate}
\end{figure}

{\bf Deterministic $T$-Gates}
Our final experiment is comparatively quite simple but highly relevant to the construction of Fault-tolerant quantum circuits \cite{PPND15}.  When designing high level quantum circuits, the universal gate library of choice is the Clifford $+T$ set.  There has been significant work in both compiling and optimising circuits of this form \cite{G06,WS14,GLRSV13,WBCB14,DSMN13,PDF16}.  In fault-tolerant models, $T$-gates are generally applied using teleportation circuits with an appropriate ancillary state that is prepared using complicated protocols such as state distillation \cite{BK05+}.  These circuits are intrinsically probabilistic, and for the $T$-gate, corrections need to involve the active application of a subsequent $S$-gate.  Hence there was the potential that high level circuit construction would need to be dynamic, adjusting itself every time a $T$-gate is applied in the hardware.  However, recent work has shown how to construct {\em deterministic circuitry} for any high-level algorithm \cite{F12+,PPND15}.  Sub-circuits known as selective source and selective destination teleportation \cite{F12+} are used to patch in correction operations for each $T$-gate via choosing to measure certain qubits in either the $X$- or $Z$ basis.  This deterministic $T$-gate is illustrated in Figure \ref{fig:Tgate}

The operations on the first two qubits are the teleported $T$-gate, which utilises a magic state ancilla, $\ket{A}=\ket{0}+e^{i\pi/8}\ket{1}$, a subsequent CNOT and $Z$-basis measurement to enact the gate.  The logical result of the $Z$-measurement determines if a $T$-gate or a $T^\dagger$-gate is applied (the $T$-gate occurs when a $\ket{0}$ result is measured).  Depending on this measurement result, a possible $S$-gate correction needs to be applied, which is a second teleportation circuit utilising a second magic state ancilla, $\ket{Y}=\ket{0}+e^{i\pi/4}\ket{1}$.  The $S$-gate correction can also require a correction, but this correction is a Pauli $Z$-gate and hence does not need to be actively applied.  

The circuit in Figure \ref{fig:Tgate} uses two circuits known as selective source and selective destination teleportation \cite{F12+} to put the $T$-gate into a form called ICM \cite{PPND15}.  An ICM form of a quantum circuit consists of a layer of qubit (I)nitialisations, an array of (C)NOT operations and a time staggered series of $X$- and $Z$-basis (M)easurements.  The choices of $X$- and $Z$-basis measurements are determined by the initial $Z$-basis measurement in the $T$-gate teleportation circuit and can dynamically {\em patch} in the correction circuit (or not).  

\begin{figure}[ht!]
\begin{center}
\resizebox{\linewidth}{!}{\includegraphics{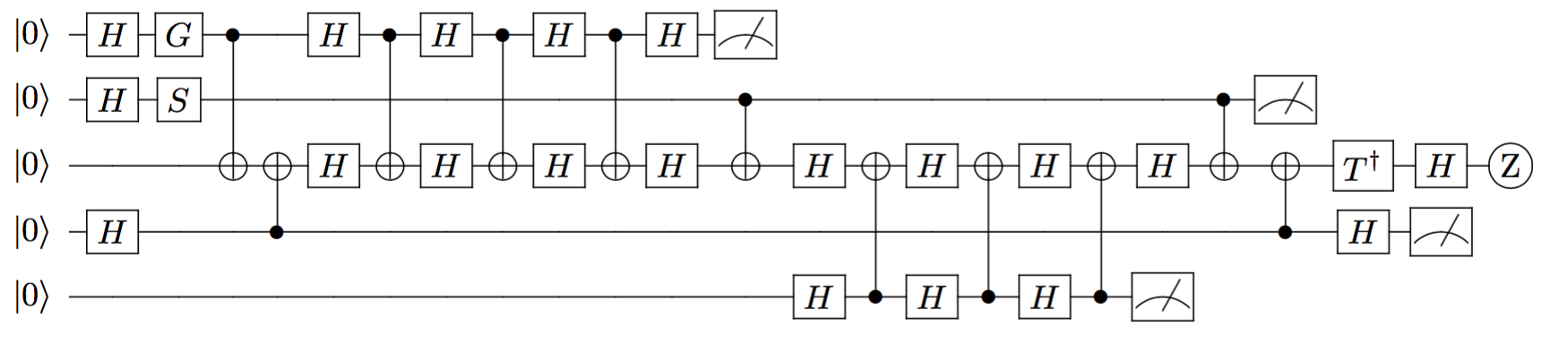}}
\end{center}
\caption{{\bf Implementation of the deterministic $T$-gate in the QE chip.}  The gate $G$ can be chosen to be either 
$T$ or $T^\dagger$ and the circuit will output the same.  SWAP operations are again needed due to the QE architecture, 
at which point the four measurements are either in the basis $Z_1X_2X_4Z_5$ or $X_1Z_2Z_4X_5$, depending on the initial 
choice for $G$.  The output qubit is rotated by $HT^\dagger$ and measured, the result should always be $\ket{0}$.}
\label{fig:Tgate2}
\end{figure}

We can simulate this circuit in the QE interface, but since the original circuit requires 6-qubits, we instead emulate the application of the $T$- or $T^\dagger$-gate directly and then based on which gate we choose, measure the four other qubits in the appropriate basis to teleport a $T$-gate to the output regardless of weather we choose $T$- or $T^\dagger$ at input.  Unfortunately while the QE interface does allow for qubit tomography to be performed, it does not allow both standard basis measurements and tomographic mapping on the same circuit run.  Therefore, to confirm the application of the $T$-gate, we simply reverse the initial circuit and confirm that $\ket{0}_{in} \rightarrow \ket{0}_{out}$.  The circuit implemented in the QE chip is shown in Figure \ref{fig:Tgate2}.  

\begin{figure*}[ht!]
\begin{center}
\resizebox{\linewidth}{!}{\includegraphics{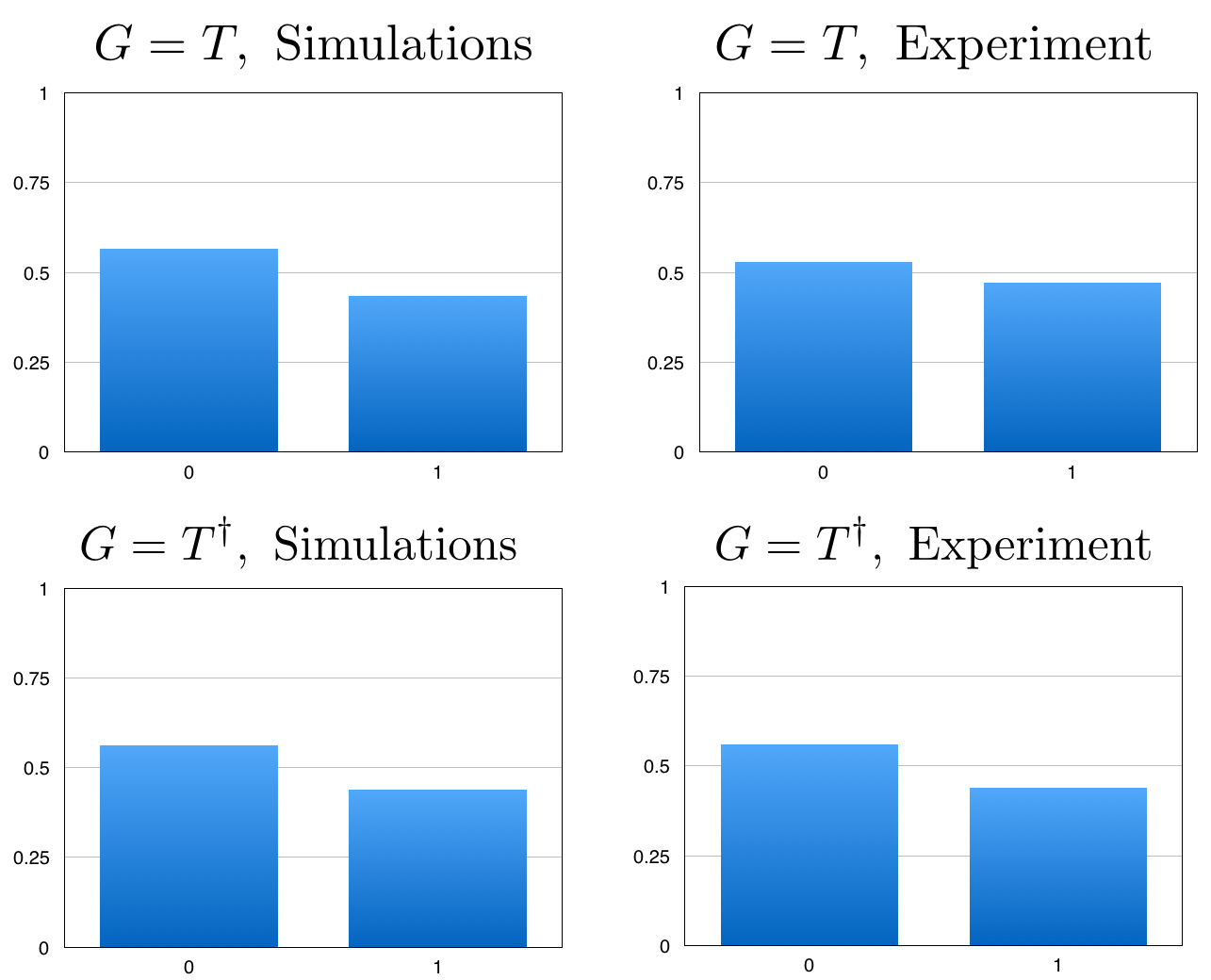}}
\end{center}
\caption{{\bf (colour online) Simulation and Experimental results for $G=\{T,T^\dagger\}$.}  For both simulations and experiments there is only a slightly higher probability of measuring he expected result ($\ket{0}$), again quantum errors for a large circuit in the QE chip is likely to blame.  However, as explained in the main text, this circuit only has a maximum probability of success of 0.75 in the ideal case.}
\label{fig:Tgateexp}
\end{figure*}

Extra SWAP operations are needed in Figure \ref{fig:Tgate2} because of the restrictions of the QE architecture.  On the top qubit we apply either the $G=T$ or $G=T^\dagger$ and dependent on that choice the four subsequent measurements are either in the $\{Z,Z,X,X\}$ basis or the $\{X,X,Z,Z\}$ basis.  Once classical Pauli corrections based on measurements are taken into account [Table \ref{Pauli}, which we will discuss more shortly] and tracked, the inverse gate $T^\dagger$ (as the output should {\em deterministically} be $T\ket{+}$ regardless of the choice for $G$) is applied and we measure the output qubit in the $X$-basis.  In the absence of circuit errors, the $X$-basis measurement should always return $\ket{0}$, indicating that the circuit dynamically applies the $S$-correction and the output is always a $T$-rotation.  Simulations and experiments with the QE interface is illustrated in Figure \ref{fig:Tgateexp}, again using 8192 instances with the SE omitted from the plot.  The requirement for us to {\em undo} the initial gates to confirm the circuit introduces interesting behaviour related to Pauli tracking \cite{PDNP++} that is highly relevant for large scale operations of a quantum computer.

\begin{table}[ht!]
\begin{tabular}{ccc}
\hline
Measurement  Basis 	& Results		& Corrections	\\
\hline
$XZZX (ZXXZ)$ & (0,0,0,0) & I (I)\\
& (1,0,0,0) & Z (X)\\
& (0,1,0,0) & X (Z)\\
& (1,1,0,0) & ZX (ZX)\\
& (0,0,0,1) & Z (X)\\
& (1,0,0,1) & I (I)\\
& (0,1,0,1) & ZX (ZX)\\
& (1,1,0,1) & X (Z)\\
& (0,0,1,0) & X (Z)\\
& (1,0,1,0) & XZ (XZ)\\
& (0,1,1,0) & I (I)\\
& (1,1,1,0) & Z (X)\\
& (0,0,1,1) & XZ (XZ)\\
& (1,0,1,1) & X (Z)\\
& (0,1,1,1) & Z (X)\\
& (1,1,1,1) & I\\
\hline
\end{tabular}
\caption{{\bf Pauli corrections for each of the four measurements in the deterministic $T$-gate.}  For each of the sixteen possible measurement results in Figure \ref{fig:Tgate2}, an appropriate Pauli correction to the output is needed.  This correction needs to be {\em known} before subsequent gates are applied (as they may need to be altered). }
\label{Pauli}
\end{table}

\begin{figure*}[ht!]
\begin{center}
\resizebox{0.75\linewidth}{!}{\includegraphics{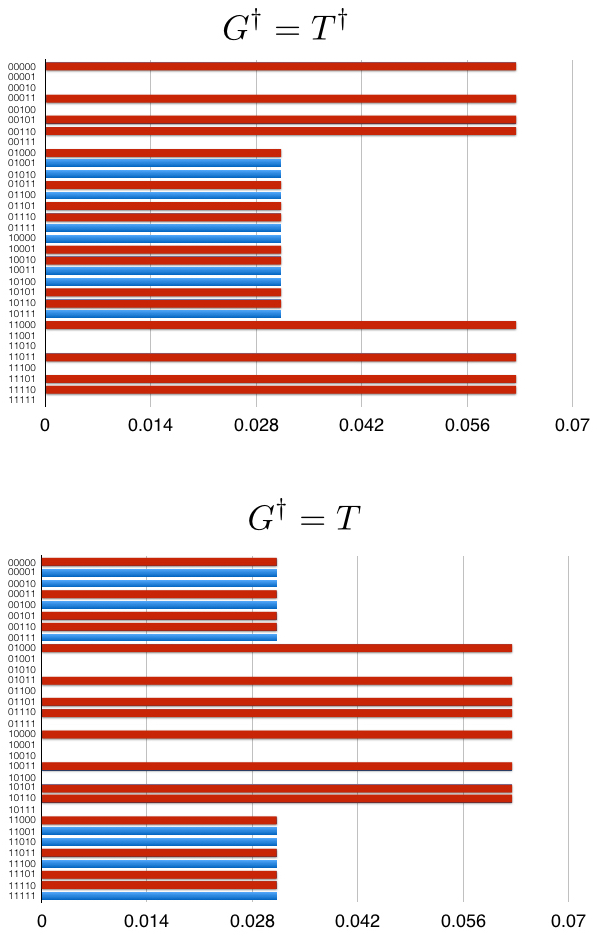}}
\end{center}
\caption{{\bf (colour online) Output distribution for an ideal application of the deterministic $T$-gate}.  We show the output distribution for an ideal 
application of the deterministic $T$-gate for $G=T$.  Because Pauli $X$-corrections arising from the teleportation circuits do not commute with the $T^\dagger$ gate used to invert the circuit and confirm output, the success probability has a maximum of 75\%.  The red bars show the binary outputs that, when corrected, are used to infer a $\ket{0}$ state output.  If we change the $T^\dagger$ gate used at the end of the circuit to $T$, the probabilities invert, since now the output will be wrong when {\em no} $X$-correction from the teleportation circuit is needed.}
\label{fig:Tgateideal}
\end{figure*}

Looking at both the simulation and experimental results, we observe the correct answer with a probability much less than one.  This is not only caused by experimental noise.  In Figure \ref{fig:Tgateideal} we illustrate the results for an ideal application of the circuit (again performed within the QE interface).  Even in ideal circumstances, the correct result is only observed (after Pauli correction) 75\% of the time.  The other 25\% is where the wrong answer is reported {\em and} $X$-corrections are required due to the teleportation operations.  This is because the $T^\dagger$-gate used to invert the circuit for verification does not commute with this corrective bit-flip gate and the QE chip does not allow us to do {\em dynamic feedforward}, i.e. change circuits based upon classical measurement result obtained earlier.  

In a real operating quantum computer, the result of the four teleportation measurements (particularly the $Z$-measurements that induce $X$-corrections) need to be known {\em before} the application of the next gate.  In this case, the second $T^\dagger$-gate.  If an $X$-correction is present, the fact that $TX = XT^{\dagger}$ implies that we would need to interchange a desired $T^\dagger$ gate with a $T$ and visa versa if a bit-flip correction is present on the input.  This bit-flip correction may come from circuits such as the deterministic $T$-gate, but they can also come from other sources such as the error-correction underlying the circuit \cite{FMMC12} or corrections coming from the tracking of information from a topological quantum circuit \cite{PDNP14+}.  

In Figure \ref{fig:Tgateideal} we can reverse the distribution of probabilities by changing the gate to we use to invert the original $T$-gate to a from $T^\dagger$ to $T$.  In this case, if $X$-corrections are not required the inverse gate is wrong and the output does not go back to $\ket{0}$.  In Figure \ref{fig:Tgateideal} we illustrate all 32 possible outputs where the {\em red} are results that after appropriate Pauli corrections and tracking, result in the output measuring $\ket{0}$.  The basis states with approximately 3\% probability of being observed have $X$-corrections from the teleportation circuits that interfere with the function of the final $HT^\dagger$ gate needed to invert the circuit.  In a circuit that would be {\em dynamically} changed dependent on measurement results, this would not occur and the output probability of $\ket{0}$ in the circuit would be 100\%. 

The results of this experiment directly demonstrate the notion of why $T$-depth is an important concept \cite{F12+}, the fact that certain results need to be known which prevents us from building all circuits with a $T$-depth of one (even through from a purely circuit perspective, this should look possible \cite{PPND15}).  If dynamic circuit changes were possible within the QE interface, we could demonstrate a fully deterministic $T$-gate.

\section{Discussion}

In all four experiments we are simply looking at the output of the IBM chip rather than examining in detail the error behaviour caused by noise in the chip, hence we have not included any significant analysis of the error properties of each qubit and/or gate (which is available from the QE interface for each run).  Analysing the detailed output from each experiment against the error data characterised for the QE chip would be interesting further work to try and refine simulation models to more accurately model how the actual device would behave given the input circuit and error characteristics of the chip.  

Even without performing rigorous error analysis on the output, in all four experiments we do see the expected results, with some experiments more susceptible to noise than others.  The simulation feature in the QE interface also shows variability depending on the experiment we conducted.  Out of the four experimental results, only the error correction Rabi experiment, showed significant deviation between the simulations and experiment.  Each of the four QE experiments allowed us to investigate the subtleties of implementing an actual quantum circuit on viable hardware and therefore required us to focus on details that were important to achieving the correct output that is often overlooked in more theoretical analysis.  The results obtained for the deterministic $T$-gate were particularly illustrative as it demonstrated the importance of having a real time, up to date Pauli frame when programming a quantum device (something that is very often overlooked).   

\section{Conclusions}
In this work, we performed four separate experiments utilising the IBM QE chip and interface, demonstrating protocols in error correction, quantum arithmetic, quantum graph theory and fault-tolerant circuit design that has not been achievable so far with an active quantum processor.  By utilising the cloud interface, these experiments could be specified, tested (in both the ideal case and with simulated errors) and then run to output noisy (but expected) results.  Each of these four protocols have direct relevance to quantum error correction, communications, algorithmic design and fault-tolerant computation and the ease of the QE interface makes subtle investigation into small quantum protocols straightforward.  

Ideally, this work will help motivate others to make use of the online hardware produced by IBM and encourage other experimental groups to make their devices accessible to other researchers for testing and theoretical development.  The five qubit quantum processor already showed significant flexibility to run tests on a large-class of quantum information protocols and increased fidelities and qubit numbers will hopefully spawn new and interesting protocols for small scale quantum computing processors.

\section{Acknowledgements}
We acknowledge support from the JSPS 
grant for challenging exploratory research and the JST ImPACT project.  We are extremely grateful to the team at IBM and the 
IBM Quantum Experience project.  This work does not reflect the views or opinions of IBM or any of its employees.   

\clearpage
\bibliographystyle{naturemag}
\bibliography{ieee.bib} 
\clearpage
\section{Supplementary material} 
Here we provide supplementary material containing the raw data for the experiments presented in the main text as well as additional experiments performed for each protocol.

\section{Error Corrected Rabi Oscillations.}

The raw data for the error correction experiment is shown below in Figure \ref{fig:ECdata} for the experiments and Figure \ref{fig:ECsims} for the simulations output from the QE interface.  Additionally Figure \ref{fig:ECsims2} presents the operating conditions for the QE chip, reported at the time of the experiment.

\begin{figure*}[ht!]
\begin{center}
\resizebox{\linewidth}{!}{\includegraphics{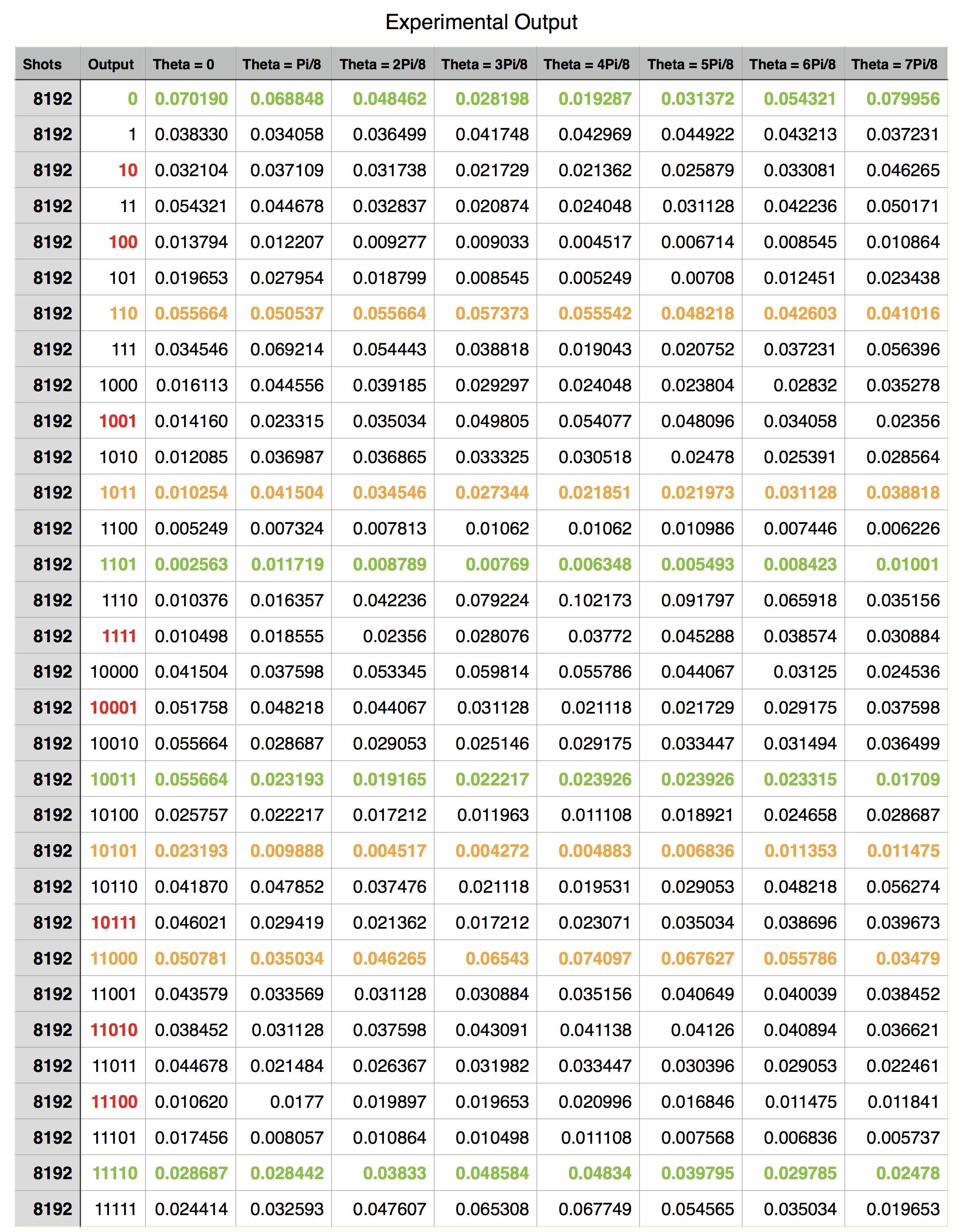}}
\end{center}
\caption{{\bf (colour online) Raw experimental data.}  Binary output for the QE chip is labeled qubits (4,2,1,5,3) for the surface code.  Red highlighted data are outputs stabilising one of the plaquette stabilizers ($Z_4Z_1Z_3$) of the surface code.  Green data satisfy trivial syndromes of both $Z$-stabilisers and correspond to a logical $\ket{0}$ while Orange data correspond to logical $\ket{1}$.}
\label{fig:ECdata}
\end{figure*}

\begin{figure*}[ht!]
\begin{center}
\resizebox{\linewidth}{!}{\includegraphics{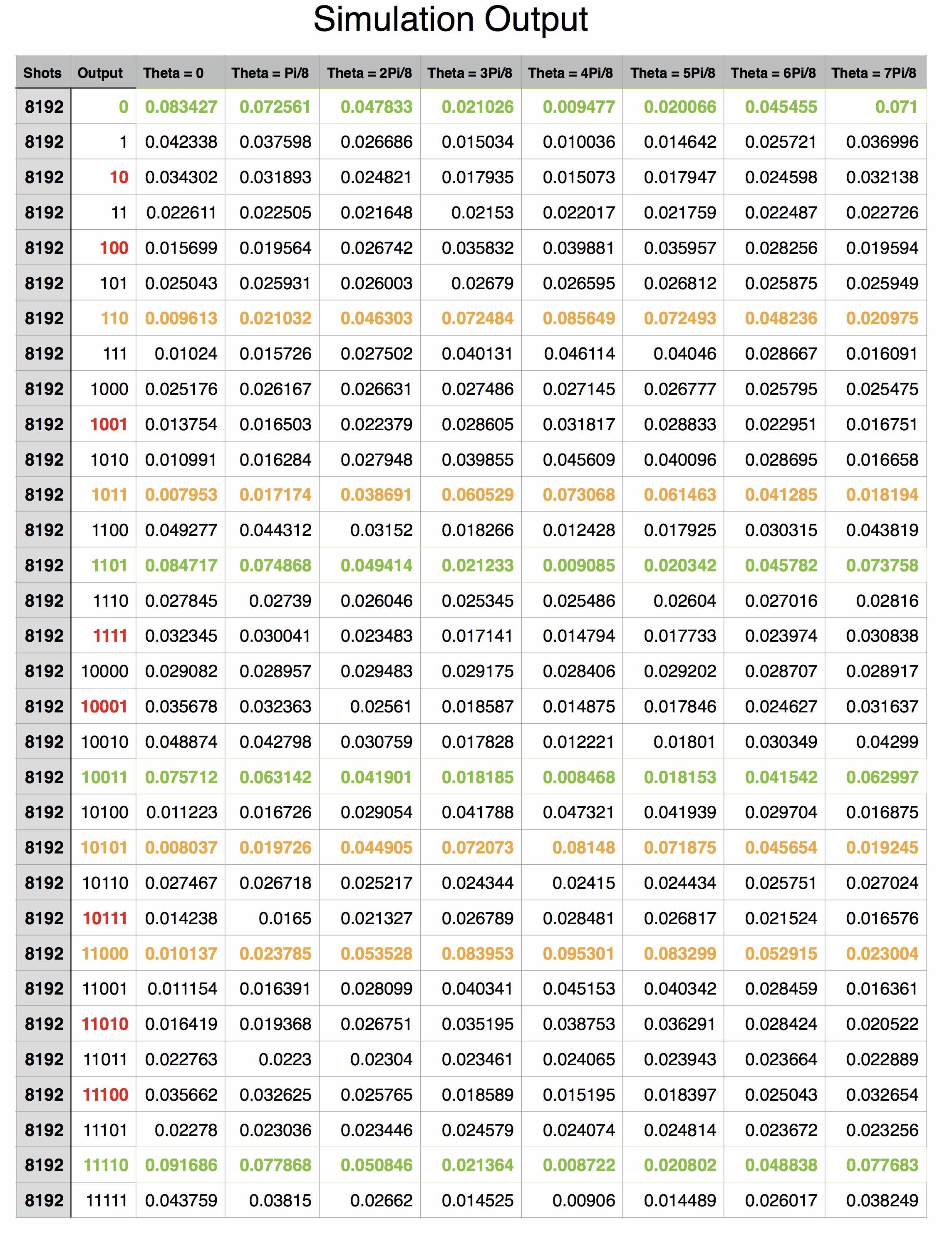}}
\end{center}
\caption{{\bf (colour online) Simulation data.}  Binary output for the QE chip is labeled qubits (4,2,1,5,3).  Red highlighted data are outputs stabilising one of the plaquette stabilizers ($Z_4Z_1Z_3$) of the surface code.  Green data satisfy trivial syndromes of both $Z$-stabilisers and correspond to a logical $\ket{0}$ while Orange data correspond to logical $\ket{1}$.}
\label{fig:ECsims}
\end{figure*}

\begin{figure}[ht!]
\begin{center}
\resizebox{\linewidth}{!}{\includegraphics{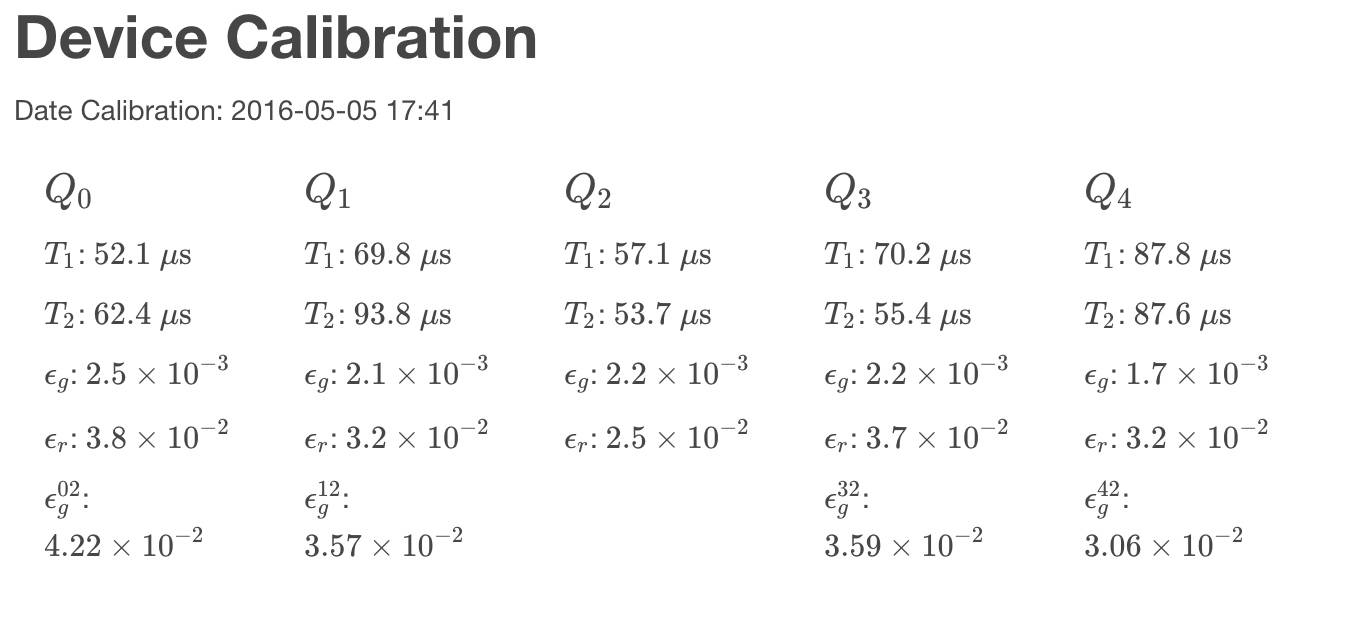}}
\end{center}
\caption{{\bf Calibration data for the Error correction experiment provided by the QE system.}}
\label{fig:ECsims2}
\end{figure}

In the main text we detailed a Rabi-oscillation experiment in the $Z$-basis.  However, the surface code is a full quantum code and therefore can correct both bit-flips and phase-flips.  To see similar results for phase errors, we need to perform a Rabi oscillation experiment in the $\ket{+,-}$ basis.  From the circuit shown in the main text, the easiest way to do this is to create the state $\alpha\ket{0} + \beta\ket{1}$, then perform a logical Hadamard and measure the logical qubit in the $\ket{+,-}$ basis.  Considering the distance two surface code is a symmetric planar code, we can perform a logical Hadamard by performing a transversal Hadamard operation on each of the five qubits and then inverting the lattice across the diagonal (which simply means swapping qubits two and four.  

However, examining Figure 2a. in the main text, after a transversal set of Hadamards, plus a second set of transversal Hadamards to realise $X$-basis measurements, the only difference in the circuits is the Swapping of qubits 2 and 4, which when you consider that the logical $X$-operator is defined as $X_2X_4$ in the lattice, the error corrected Rabi experiment in the $\ket{+,-}$ basis is exactly the same circuit and interpretation of measurement results.

Additionally, if you take the effort to examine the circuit necessary to directly inject the $\alpha \ket{+}+\beta\ket{-}$ state into the code, you will see that again the circuit is identical (up to a relabelling of qubits 2 and 4).  Hence a direct experiment for a Rabi Oscillation in the $\ket{+,-}$ basis will produce identical results to those already shown.  

If the IBM QE interface can be expanded in the future to allow for the construction of arbitrary rotations (either directly with additional gates, or by giving more circuit space and fidelity to construct arbitrary rotations out of a Clifford $+T$ gate set, the possibility of doing Rabi experiments in rotated bases will be possible.  We will leave that up to future work. 

\section{Fourier Addition}

In the main text we only provided a small subset of all possible addition experiments.  The following figures illustrate the rest of the binary inputs that could be used, and more complex superposition inputs.  The correct results are indicated in red in each plot.  Not all experiments return the correct results.  The calibration data during these runs is illustrated in Figure \ref{fig:addercal}
\begin{figure*}[ht!]
\begin{center}
\resizebox{0.8\linewidth}{!}{\includegraphics{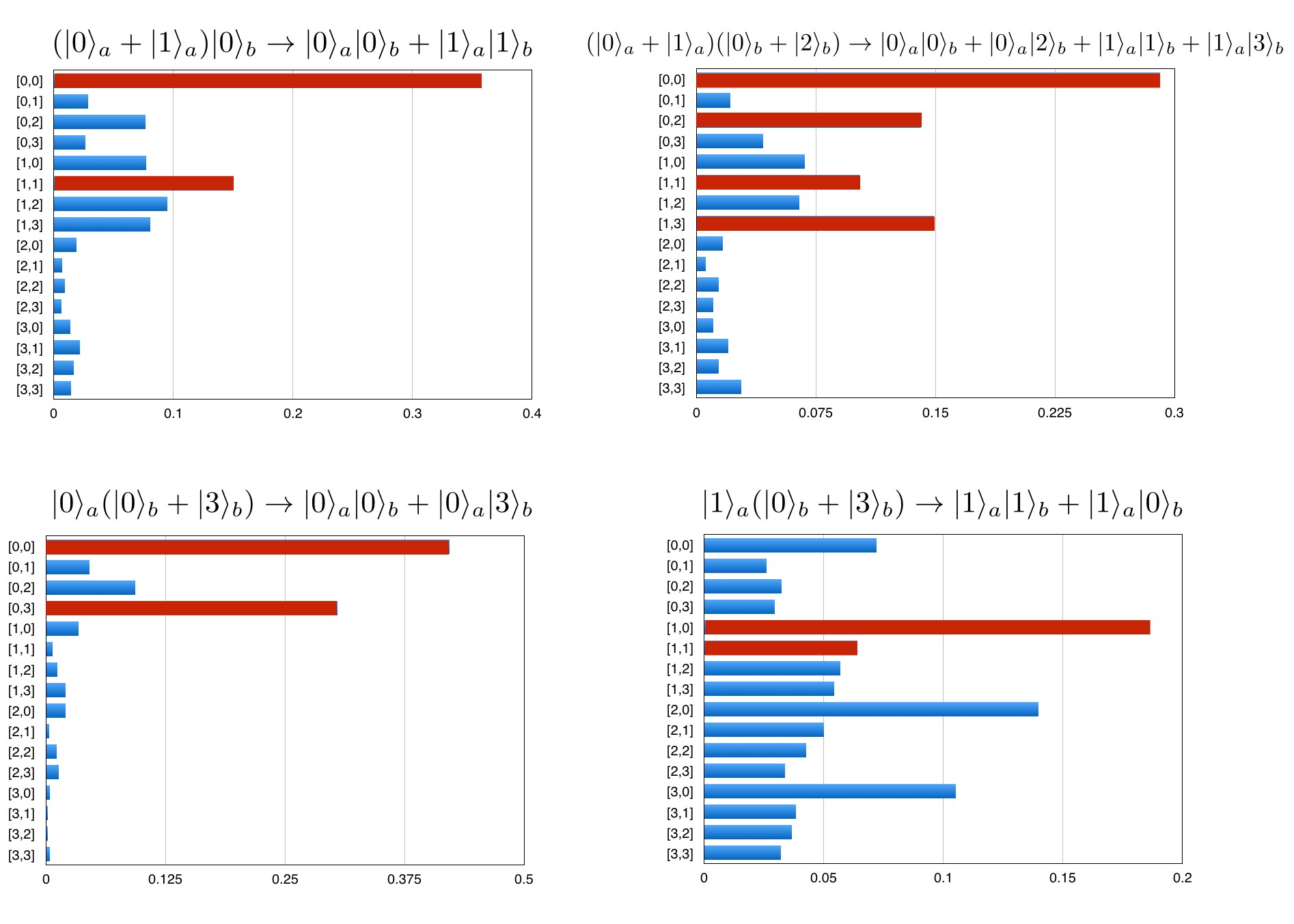}}
\end{center}
\caption{{\bf (colour online) Experimental Fourier Addition.}.  Correct mappings are illustrated above and the correct results are highlighted in red.}
\label{fig:adderex4}
\end{figure*}

\begin{figure*}[ht!]
\begin{center}
\resizebox{0.8\linewidth}{!}{\includegraphics{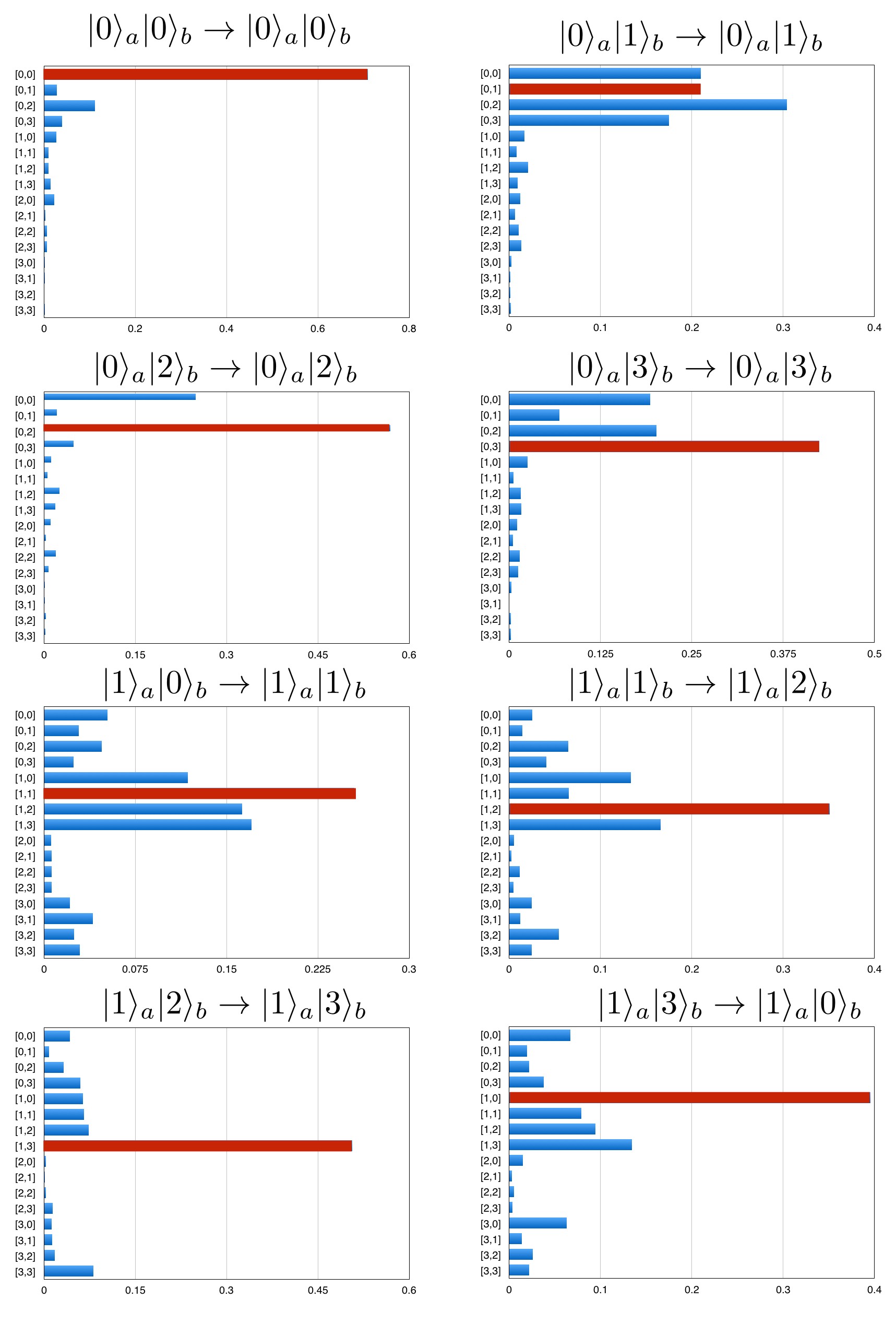}}
\end{center}
\caption{{\bf (colour online) Experimental Fourier Addition.}.  Correct mappings are illustrated above and the correct results are highlighted in red.}
\label{fig:adderex1}
\end{figure*}

\begin{figure*}[ht!]
\begin{center}
\resizebox{0.8\linewidth}{!}{\includegraphics{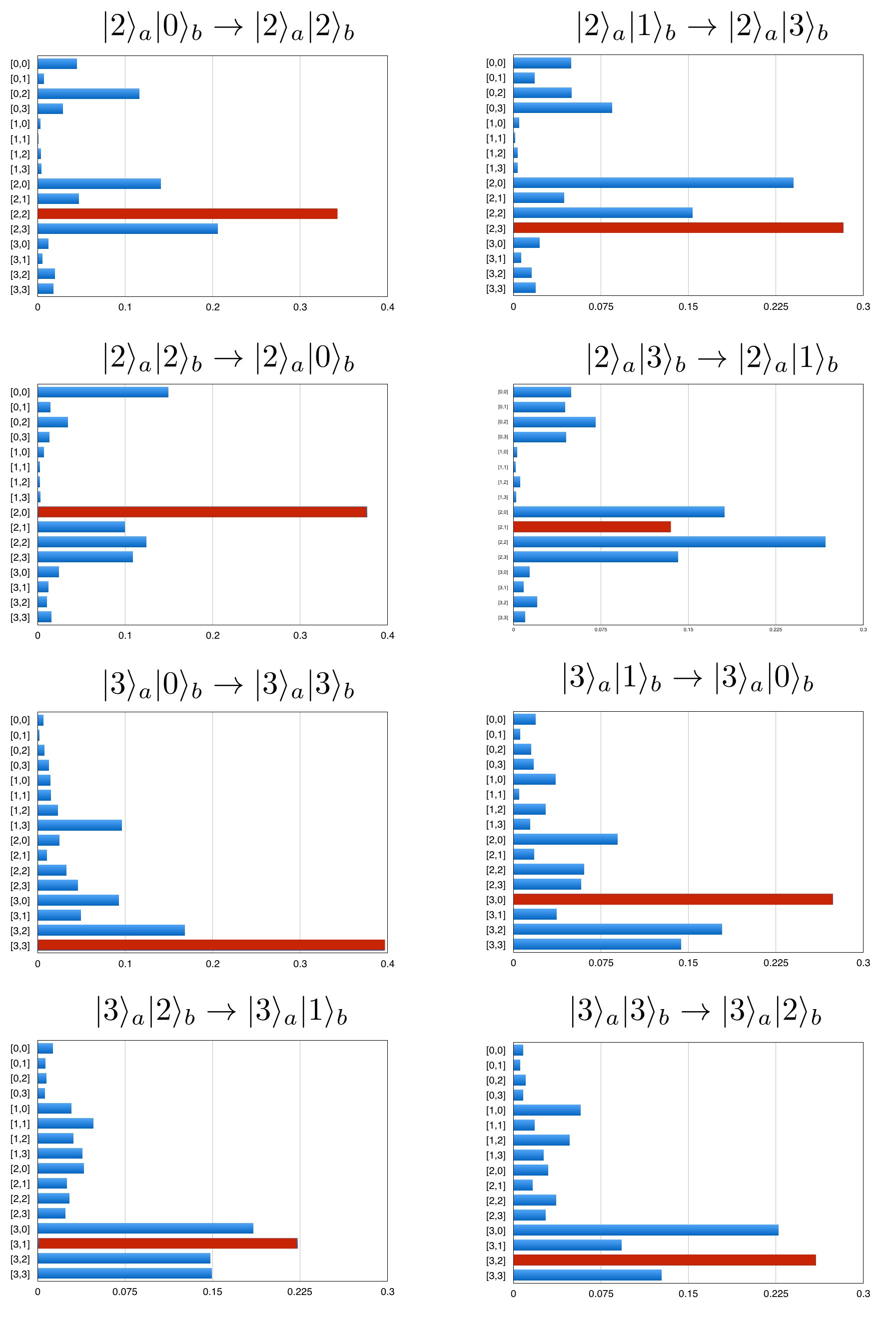}}
\end{center}
\caption{{\bf (colour online) Experimental Fourier Addition.}.  Correct mappings are illustrated above and the correct results are highlighted in red.}
\label{fig:adderex2}
\end{figure*}

\begin{figure*}[ht!]
\begin{center}
\resizebox{0.8\linewidth}{!}{\includegraphics{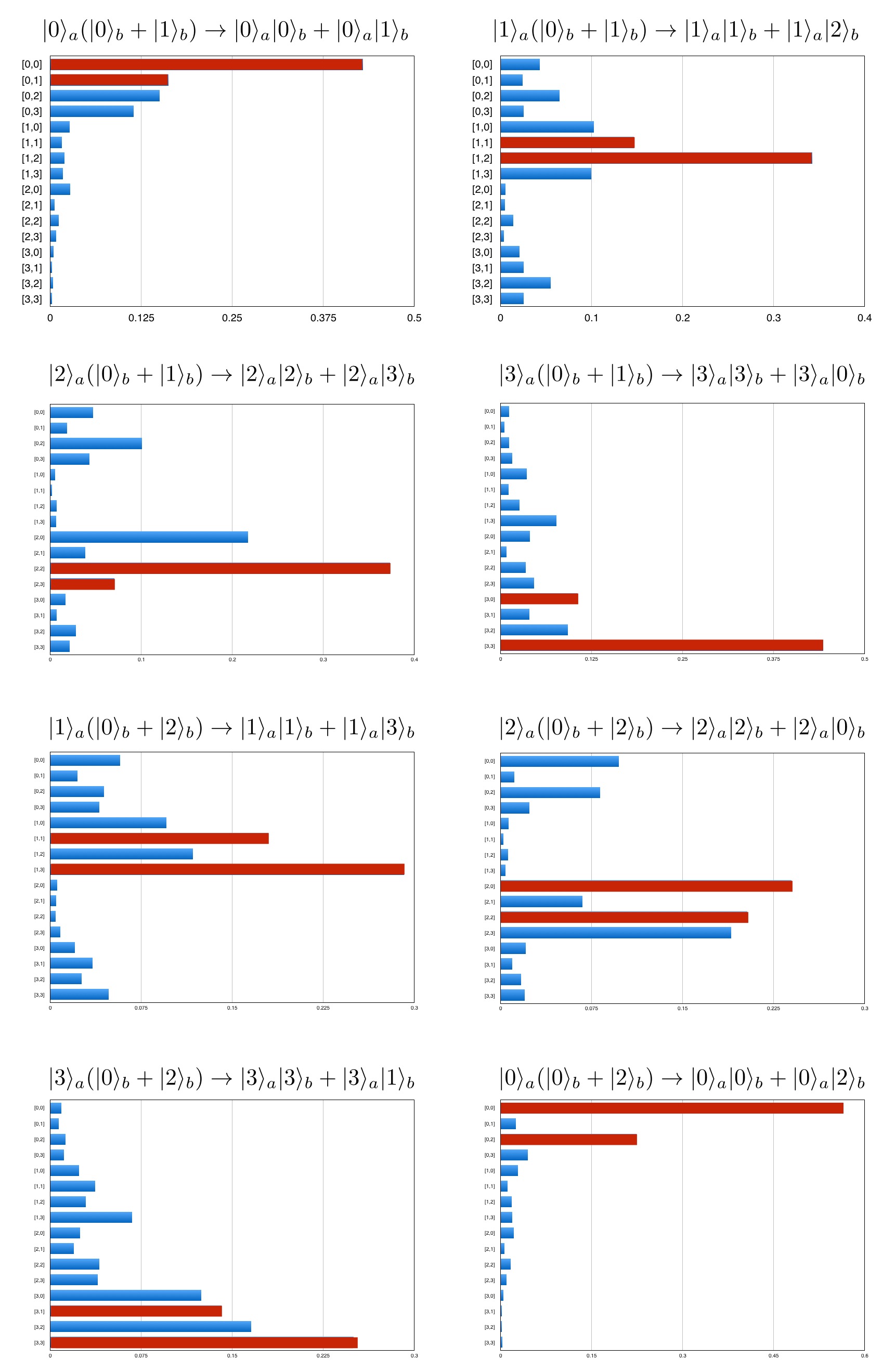}}
\end{center}
\caption{{\bf (colour online) Experimental Fourier Addition.}.  Correct mappings are illustrated above and the correct results are highlighted in red.}
\label{fig:adderex3}
\end{figure*}

\begin{figure*}[ht!]
\begin{center}
\resizebox{0.8\linewidth}{!}{\includegraphics{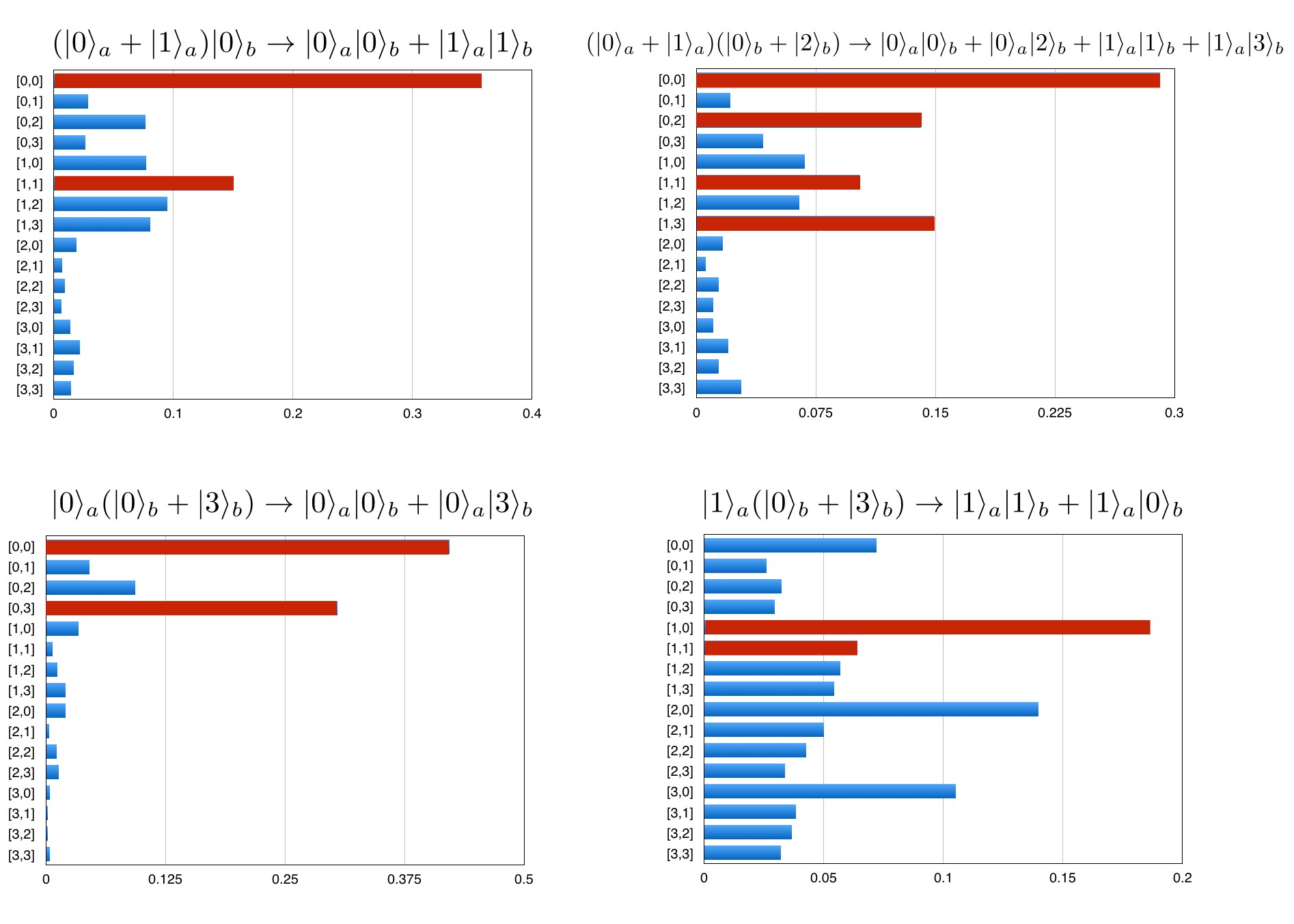}}
\end{center}
\caption{{\bf (colour online) Experimental Fourier Addition.}.  Correct mappings are illustrated above and the correct results are highlighted in red.}
\label{fig:adderex4}
\end{figure*}

\begin{figure*}[ht!]
\begin{center}
\resizebox{0.8\linewidth}{!}{\includegraphics{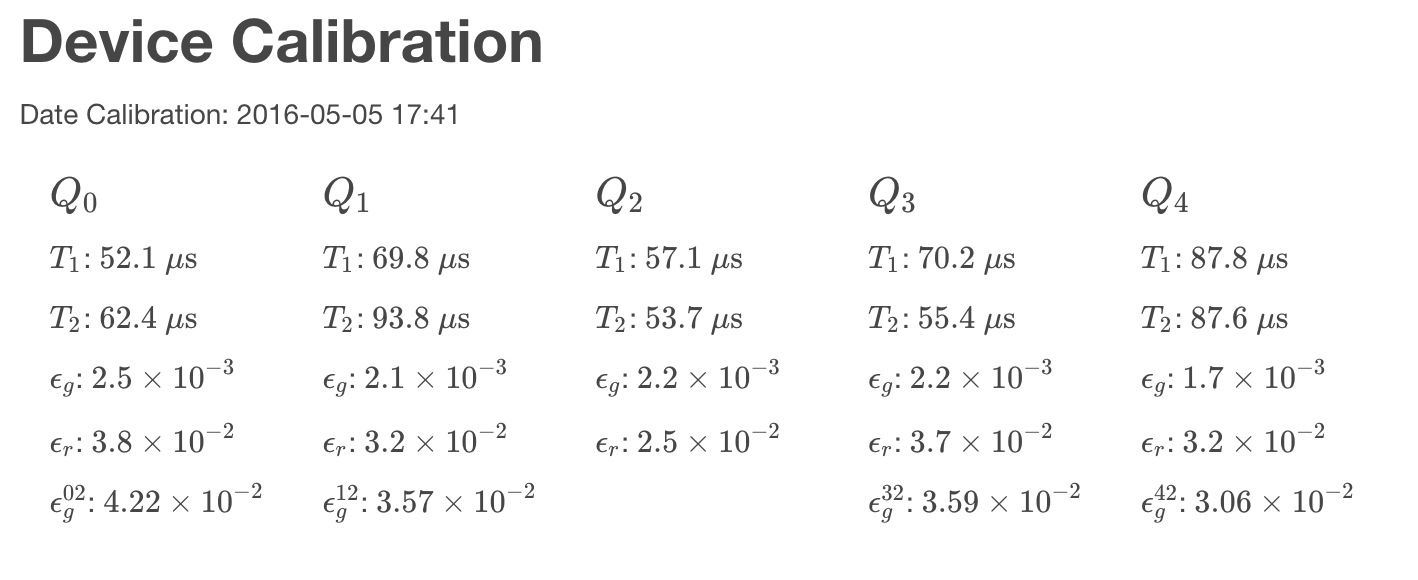}}
\end{center}
\caption{{\bf IBM Calibration data during the Fourier Addition Runs}.}
\label{fig:addercal}
\end{figure*}

Most of the Fourer addition circuits used did return the correct results a plurality of the time.  But there were a few examples where only half of the wavefunction was the dominant result, with {\em non answers} becoming the second most probable.  We did not include the QE simulations in these results as the rough overlap in the main text was significantly better than the error corrected Rabi oscillations and showed effectively the same behaviour.  

\section{Quantum Graph Complementarity}  
In the main text we illustrated experiments looking at the conversion of one type of quantum graph state to another.  The main text included an example of converting a star graph (GHZ state) to a completely connected graph state and then back again.  For a five qubit graph state, there are a large class of non-locally-equivalent graphs that each form orbital groups.  We will not attempt to simulate all of them (and we won't present here a complete orbital group).  Instead, we will focus on two permutations of a five qubit loop graph.  

\begin{figure*}[ht!]
\begin{center}
\resizebox{0.75\linewidth}{!}{\includegraphics{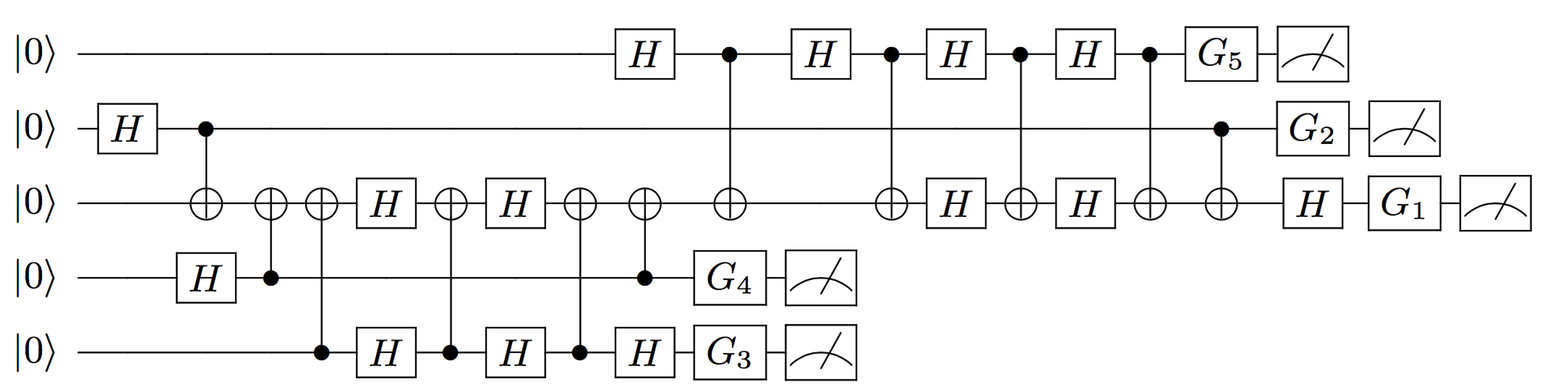}}
\end{center}
\caption{{\bf QE circuit for complementarity experiment.}  The gates, $G_i$ $i\in \{1,..,5\}$ represent local rotations to perform graph permutations or change measurement basis.  After the circuit, the qubits are re-ordered to (5,2,1,4,3).}
\label{fig:loopcir}
\end{figure*}

\begin{figure*}[ht!]
\begin{center}
\resizebox{0.75\linewidth}{!}{\includegraphics{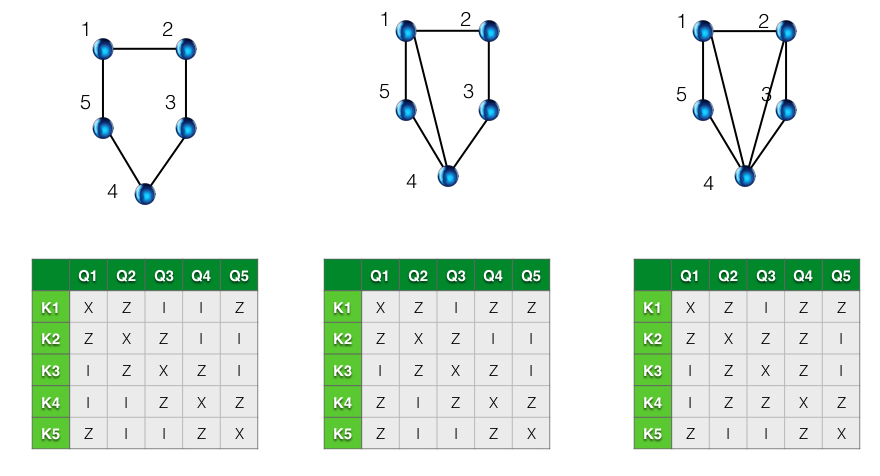}}
\end{center}
\caption{{\bf (colour online) An initial five qubit loop graph and two orbital permutations}.  Along with the structure for each graph, 
we include the stabiliser table}
\label{fig:loop}
\end{figure*}

The three graphs are illustrated in Figure \ref{fig:loop}, along with their respective stabilisers.  The two permutations were first applying the operator $G = \sqrt{X_5}\sqrt{Z_1}\sqrt{Z_4}$ and the second was obtained using by applying the operator $G=\sqrt{X_3}\sqrt{Z_2}\sqrt{Z_4}$.  The quantum circuit used in the QE interface to build the initial loop graph is illustrated in  Figure \ref{fig:loopcir}.

\begin{figure*}[ht!]
\begin{center}
\resizebox{0.75\linewidth}{!}{\includegraphics{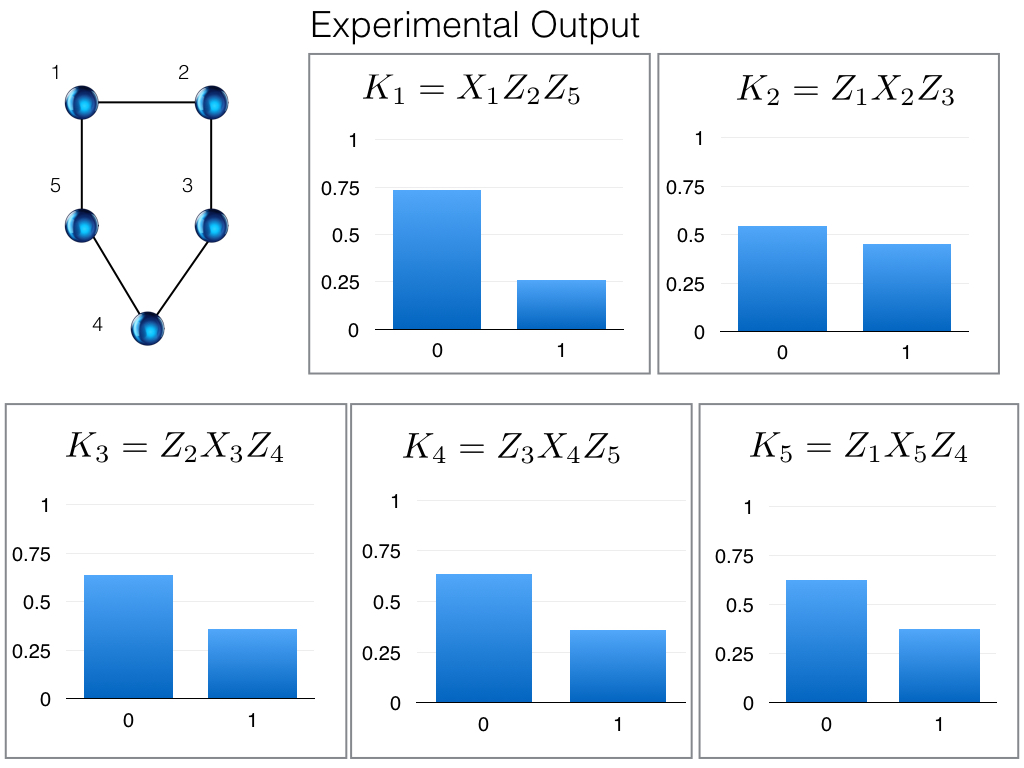}}
\end{center}
\caption{{\bf (colour online) Stabiliser measurements for the initial five-qubit ring cluster.}}
\label{fig:loopexp1}
\end{figure*}

\begin{figure*}[ht!]
\begin{center}
\resizebox{0.75\linewidth}{!}{\includegraphics{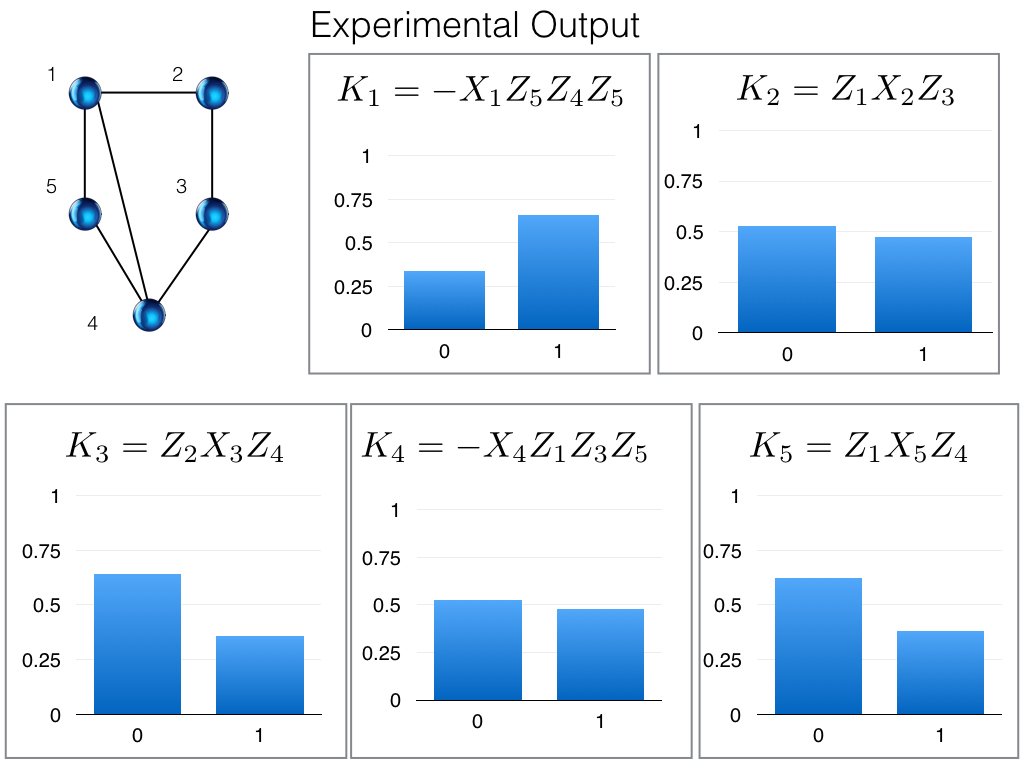}}
\end{center}
\caption{{\bf (colour online) Stabiliser measurements for the first graph permutation.}  Notice that the incorrect result is obtained for $K_4$.  The experiment was run multiple times, but most likely due to quantum noise, a near 50/50 mixture was returned.}
\label{fig:loopexp2}
\end{figure*}

\begin{figure*}[ht!]
\begin{center}
\resizebox{0.75\linewidth}{!}{\includegraphics{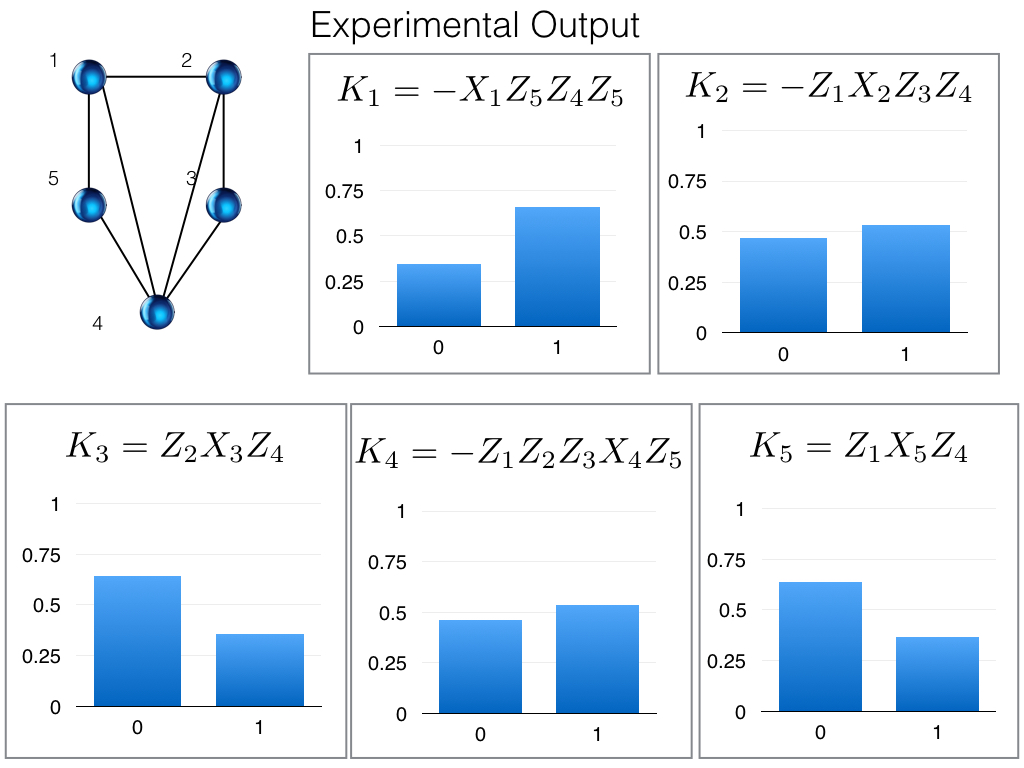}}
\end{center}
\caption{{\bf (colour online) Stabiliser measurements for the second graph permutation.}}
\label{fig:loopexp3}
\end{figure*}

The experimental measurements of each stabiliser, for each graph, are illustrated in Figures \ref{fig:loopexp1}, \ref{fig:loopexp2} and \ref{fig:loopexp3}.  Due to the increased circuit complexity to build the initial graph, stabiliser measurements suffer from much more noise than the example in the main text.  For the $K_4$ stabiliser for the first permutation, the correct output is not observed.  


\begin{figure*}[ht!]
\begin{center}
\resizebox{0.75\linewidth}{!}{\includegraphics{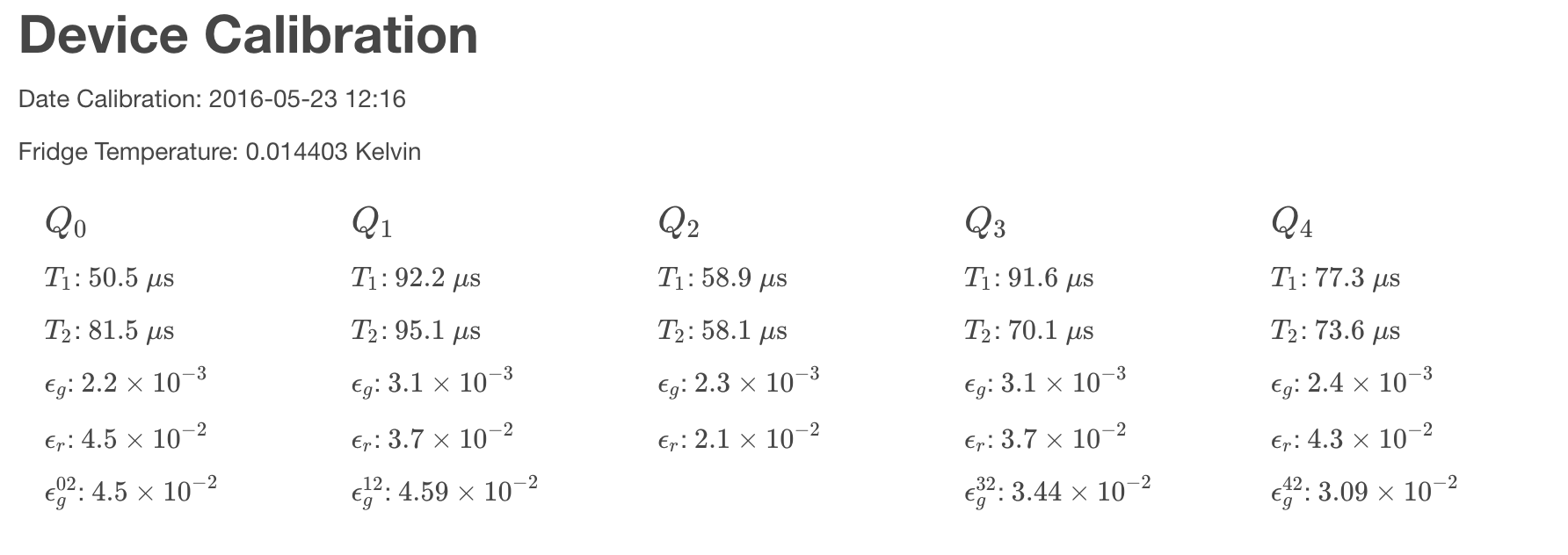}}
\end{center}
\caption{{\bf (colour online) Calibration data from the IBM machine for the second complementarity experiment.}}
\label{fig:cal2}
\end{figure*}

\begin{figure*}[ht!]
\begin{center}
\resizebox{0.75\linewidth}{!}{\includegraphics{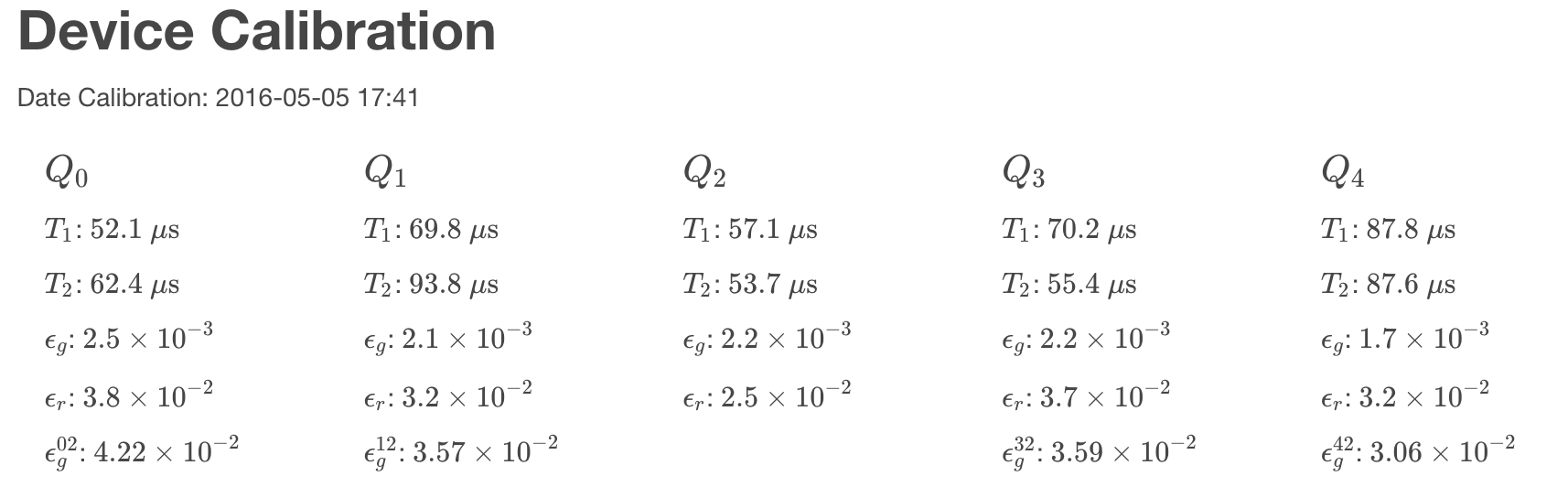}}
\end{center}
\caption{{\bf Calibration data from the IBM machine for the graph complementarity experiments in the main text.}}
\label{fig:cal3}
\end{figure*}

The IBM calibration data for this second graph complementarity experiment is illustrated in Figure \ref{fig:cal2}, while calibration data for the star graph experiment illustrated in the main text is illustrated in Figure \ref{fig:cal3}.

\section{Deterministic $T$-gate}

\begin{figure*}[ht!]
\begin{center}
\resizebox{0.75\linewidth}{!}{\includegraphics{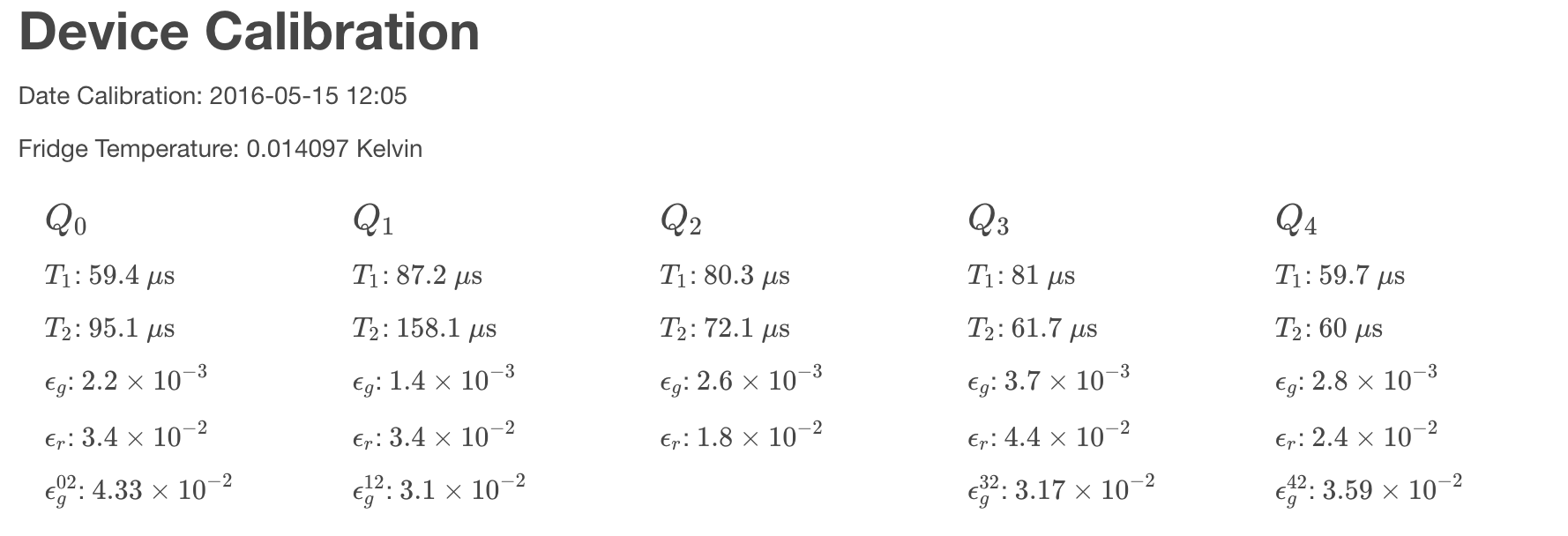}}
\end{center}
\caption{{\bf Calibration data from the IBM machine for the deterministic $T$-gate experiment in the main text.}}
\label{fig:cal4}
\end{figure*}

We have chosen not to elaborate further on this experiment.  In this section we simply provide the IBM calibration data [Figure \ref{fig:cal4}] for the results in the main text.

\end{document}